\documentclass{article}
\usepackage[margin=1in]{geometry}  

\usepackage{soul}
\usepackage{microtype}
\usepackage{graphicx}
\usepackage{subcaption}
\usepackage{booktabs} 
\usepackage{bbm}
\usepackage{dsfont}
\usepackage{authblk}  

\usepackage{hyperref}
\usepackage{amsmath,amssymb,amsfonts}
\usepackage{amsthm}
\usepackage{xcolor}
\usepackage{algorithmic}
\usepackage{cite}
\usepackage{textcomp}
\usepackage{listings}
\usepackage{booktabs}
\usepackage{array}
\usepackage{xspace}
\usepackage{pifont}
\usepackage{multirow}
\usepackage{mathtools}
\usepackage{thmtools}
\usepackage{natbib}
\usepackage{enumitem}
\usepackage[nameinlink]{cleveref}
\newcommand{\cmark}{\ding{51}}
\newcommand{\xmark}{\ding{55}}

\newcommand{\diff}[1]{{\bfseries\underline{#1}}}

\newcommand{\tool}{{\scshape \bfseries Monty}\xspace}

\newcommand{\mypara}[1]{\emph{\textbf{#1}.\ }}

\newcommand{\Ll}{\mathcal{L}}

\newtheorem{definition}{Definition}

\newcommand{\ignore}[1]{}

\title{Faithful Autoformalization of Natural Language Assertions}

\author[1]{Hongyi Liu}
\author[2]{Madhusudan Parthasarathy}
\author[1]{Adithya Murali}

\affil[1]{University of Wisconsin-Madison}
\affil[2]{University of Illinois at Urbana-Champaign}

\affil[1]{\texttt{\{adithyamurali, hliu794\}@cs.wisc.edu}}
\affil[2]{\texttt{madhu@illinois.edu}}

\date{}

\begin{document}

\maketitle

\begin{abstract}

Formal contracts are essential for software testing and verification, yet writing them remains labor-intensive and error-prone. LLMs offer a promising path toward autoformalization: synthesizing executable assertions from natural-language specifications and thereby bridging the gap between informal developer intent and formal executable specifications. We present \tool: an autoformalization framework for assertions that tackles the challenges of expectations of validity of assertions and ambiguity in natural-language.
Our techniques are based on filtering formalizations using a novel conformance score metric and validity scores obtained from testing the code against formalized assertions. We evaluate our approach on 541 assertion-generation tasks derived from 22 collection-like Java classes, and show that our technique produces the ground truth more reliably (improving upto 20 points in precision on average) than when using LLMs naively to translate assertions.

\end{abstract}

%
%
%

\section{Introduction}
\label{sec:introduction}

Formal specifications for modular code, in terms of contracts or postcondition assertions, have been a cornerstone in enabling robust and effective software development. Formal specifications realized as runtime checks enable testing code at the unit level, enable formal verification efforts, and enable evolution of software where modules can be swapped with others that satisfy the same specifications. The paradigm of Design by Contract~\cite{dbc} concretizes this approach to development of software, and several specification languages for contracts have been developed over decades, ranging from Eiffel~\cite{meyer1992eiffel}, JML~\cite{Leavens1999JMLAN}, Spec\#~\cite{specsharp}, and Code Contracts~\cite{codecontracts}, the latter two at Microsoft. However, despite the advantages, we have always struggled in getting programmers to write formal specifications, primarily due to it being perceived as a ``burden'' that slows the time to ship out software.

The advent of AI opens up two avenues that transform the framework of writing code with contracts. First, AI opens up the exciting possibility of \emph{natural language specifications}  that are then autoformalized into formal and executable contracts (executable logics).  This considerably lowers the burden of writing formal specifications, and is akin to current documentation that is used for understanding code between human software developers.
Second, AI-based automated coding/programming has made great advances in recent years, allowing programmers to simply state the intent of their programming task. This however creates a large gap between the intent of the programmer and the code~\cite{Lahiri}. Formal specifications can help bridge this gap--- we can have developers write natural-language specifications at the level of modules, and have autoformalization of these assertions used to check the AI-written code using testing, or even formally prove the code correct against derived formal specifications.
This general framework sits well in \emph{specification-driven development using AI}~\cite{kiro,githubspeckit}, where structured, human-readable specifications are authored and maintained during AI-written code development. 

The key problem that enables both the above applications is the problem of \emph{reliable autoformalization of assertions}.
Reliability is important--- when the tool succeeds, we need to be able to trust that the formalizations are faithful translations of the natural-language specifications.
More formally, in addition to the usual accuracy metric, we are interested in high \emph{precision} of the translation framework, i.e., in reducing the number of incorrect translations.
We identify and address two key challenges for autoformalizing assertions: (a) how do we effectively assess the faithfulness of the formalization, (b) how do we resolve the inherent ambiguity in specifications written in natural language?

While the former problem is a general problem in autoformalization, the latter has unique aspects in the setting of translating assertions. In particular, ambiguities in natural-language assertions are particularly troublesome when 
they interpret a valid assertion as an invalid one (raising false positives that lead to doubting the code is correct) or interpret invalid assertions as valid assertions (leading to the assertion not checking the user intent, and hence passing buggy code). Consequently, in our work, we propose to return both the most likely valid and invalid formalizations of the natural-language assertion. Furthermore, when reasonably likely valid and invalid formalizations exist, we propose active learning algorithms to query the user further in order to disambiguate the assertion.

In contrast, several existing works on autoformalization of specifications in the literature instead make the implicit assumption that the assertion is expected to be \emph{valid}~\cite{hahn2022formalspecificationsnaturallanguage, ma2025specgenautomatedgenerationformal, wen2024autospec}. While it is tempting to make this assumption in order to disambiguate the meaning of an assertion, these techniques are not useful in a context where there is no such expectation on the assertion being valid. For example, when the programmer is writing assertions to \emph{test} the code, clearly such an assumption of validity is counterproductive. Also, in the context of using assertions to validate AI-written code, it is  crucial to not assume the assertion is valid on that code!

We need a technique to check whether assertions are valid or invalid, and a key idea in this paper is to \emph{analyze the code to vet the autoformalization}. More precisely, we analyze the behavior of the program on automatically generated test inputs to determine the validity/invalidity of assertions. 


In this paper, we propose a framework of autoformalization using a combination of LLMs and program testing that (a) evaluates the confidence of various formalizations of assertions using a novel technique called \emph{clausal coverage}, (b) generates both the most likely valid formalization and the most likely invalid formalization, using metrics that include the clausal coverage metric, and (c) if necessary, disambiguates between the most likely valid and invalid formalizations using active learning by querying the programmer. 

The clausal coverage technique is a key technical contribution of our work. We translate the formal specification back to natural language, and then ask LLMs whether  every clause in this NL specification is matched by a clause of the original specification, and vice versa. We ask the LLM to provide a conformance score that captures how well the clauses match, and utilize it to estimate how robust the formalization is.

We emphasize that our work caters to autoformalizing assertions written at the level of modules/code. In the context of AI-written code, our proposed tool helps programmers write module-level specifications (in addition to overall intent) that are then vetted against the code using autoformalization and testing. We do not attempt to solve several closely related problems studied in the literature such as (a) formalizing higher application-level intent of a programmer~\cite{Lahiri}, or (b) mining formal specifications from documentation in natural language~\cite{zhong2009inferringresourcespec, pandita2012inferringmethodspec}. Our solutions, in particular the clausal coverage technique and active learning, rely on the fact that localized natural-language assertions are being formalized. 

We target formalizing contracts/assertions for a rich class of specifications of object-oriented code that encapsulates data and provides interface methods. The methods include those 
that have effects as well as those that are pure \emph{observer methods} that return information of the underlying object. Our formal specifications are based on writing logical assertions involving the rich class of observer methods on objects. The logics we target are expressive, involving Boolean combinations, bounded quantification, and multiple sorts (objects, integers, etc.). We target realistic specifications of such modules.

\paragraph*{Benchmarking, Implementation, and Evaluation} We curate several benchmark suites containing NL specifications and corresponding ground-truth formal specifications. The datasets span both synthetically generated NL specifications and manually written ones, and contain both valid and invalid assertions. Together, the datasets contain a total of 541 (NL spec, formal spec) pairs across 22 Java classes that represent collection-like data structures. We implement our approach in a tool called \tool and evaluate different research questions.

Our experiments show that \tool is effective at autoformalizing NL specifications. In particular, our approach significantly improves the \emph{precision} of autoformalization beyond simply using an LLM, i.e., how likely is it that a generated response is correct? As may be expected, the effect is more pronounced with smaller models, and \tool used with Qwen2.5-Coder (a 32B parameter model, considered relatively small) improves the precision from 75\% to 91.6\% on one dataset, and from 64\% to 85\% on another dataset. We also show that the improvement in precision is obtained while still maintaining a high recall, which shows that \tool improves the reliability of autoformalization without affecting the availability of correct responses from raw LLM translations. Finally, we also perform several ablations relating to the choice of backbone model, the conformance checking approach we contribute, and the active learning. The most interesting observation among these is that our clausal coverage method appears to be the best approach for conformance checking, outperforming baseline approaches similar to those found in related literature.

\section{Illustrative Example}
\label{sec:example}

We illustrate our pipeline with an assertion for a method in the standard \texttt{ArrayList} Java class~\cite{java21arraylist}. The \texttt{add(int index, E element)} method inserts an element at a specified position, shifts the existing suffix to the right, and then increments the size of the list. One natural-language specification for this method is:

\begin{quote}
\emph{``If the index is valid, the value at the index after insertion will be the inserted element.''}
\end{quote}

Figure~\ref{fig:arraylist-add-task} shows the skeleton of the \texttt{ArrayList} class that we provide as an input to our pipeline. Notice that the skeleton also contains the stubs of methods other than \texttt{add}. These are \emph{observer} methods, which are side effect-free functions that serve as atoms for the formal specification vocabulary (see Section~\ref{sec:prelim} for a more formal presentation). In this case, the observer methods are \texttt{get}, which returns the value at a given index, and \texttt{size}, which returns the size of the array list. The skeleton includes documentation strings and method signatures for the observer methods as well as the target method signature. The comment marked with \texttt{@@@} denotes the natural-language assertion to be formalized.


Although the NL assertion appears straightforward, a faithful formalization must avoid many pitfalls, both in terms of natural language understanding and writing semantically correct assertions for the underlying programming language. We illustrate these challenges by walking through the different stages of our solution architecture below. 

\begin{figure}[t]
\centering
\begin{minipage}{0.72\textwidth}
\begin{lstlisting}[
  language=Java,
  basicstyle=\ttfamily\footnotesize,
  breaklines=true,
  columns=fullflexible
]
class ArrayList<E> {
  /** Returns the element at the specified position. */
  E get(int index);

  /** Returns the number of elements in this list. */
  int size();

  /**
   * Inserts element at the specified position in this list.
   * Shifts the element currently at that position (if any)
   * and any subsequent elements to the right.
   *
   * @param index index at which element is inserted
   * @param element element to be inserted
   * @throws IndexOutOfBoundsException {@inheritDoc}
   */
  void add(int index, E element) {
    ...
    // @@@ If the index is valid, the value at the index after insertion will be the inserted element.
  }
  ...
}
\end{lstlisting}
\end{minipage}

\vspace{-0.8em}
\caption{Example assertion-generation task for \texttt{ArrayList.add}.}
\label{fig:arraylist-add-task}
\end{figure}

\begin{table}[t]
\centering
\footnotesize
\setlength{\tabcolsep}{4pt}
\renewcommand{\arraystretch}{1.15}

\begin{tabular}{p{0.05\textwidth} p{0.80\textwidth}}
\toprule
\textbf{\#} & \textbf{Assertion Candidates} \\
\midrule

0 &
\begin{tabular}[t]{@{}l@{}}
{\ttfamily assert (0 <= index \&\& index <= \textbackslash old(this.size()))} \\
\hspace{1.5em}
{\ttfamily => element == null \&\& this.get(index) == null
|| this.get(index).equals(element);}
\end{tabular}
\\[0.45em]

1 &
\begin{tabular}[t]{@{}l@{}}
{\ttfamily assert (index >= 0 \&\& index <= \textbackslash old(this.size()))} \\
\hspace{1.5em}
{\ttfamily => \diff{this.get(index).equals(element);}}
\end{tabular}
\\[0.45em]

2 &
\begin{tabular}[t]{@{}l@{}}
{\ttfamily assert (0 <= index \&\& index <= \diff{this.size()})} \\
\hspace{1.5em}
{\ttfamily => element == null \&\& this.get(index) == null
|| this.get(index).equals(element);}
\end{tabular}
\\[0.45em]

3 &
\begin{tabular}[t]{@{}l@{}}
{\ttfamily assert (index >= 0 \&\& index \diff{<} \textbackslash old(this.size()))} \\
\hspace{1.5em}
{\ttfamily => element == null \&\& this.get(index) == null
|| this.get(index).equals(element);}
\end{tabular}
\\[0.45em]

4 &
\begin{tabular}[t]{@{}l@{}}
{\ttfamily assert (index >= 0 \&\& index <= \textbackslash old(this.size()))\diff{)}} \\
\hspace{1.5em}
{\ttfamily => element == null \&\& this.get(index) == null
|| this.get(index).equals(element);}
\end{tabular}
\\

\bottomrule
\end{tabular}

\caption{Representative LLM-generated assertions for the \texttt{ArrayList.add} example. Candidate~\#0 is equivalent to the ground truth. Underlined text marks fragments that differ from the ground truth or cause validation failures.}
\label{tab:arraylist-add-candidates}
\end{table}

\subsection*{Generating Candidate Formalizations using LLMs}

Given the NL specification and the code skeleton above, we first use an LLM to generate candidate formal specifications in JML (Java Modeling Language)\cite{Leavens1999JMLAN}\footnote{We make some minor cosmetic adjustments to ensure we can reliably prompt an LLM to write syntactically correct assertions, and parse the generated expressions into standard JML}. Let us say that the LLM produces the five candidate formalizations shown in Table~\ref{tab:arraylist-add-candidates}. These assertions are in fact representative of the different scenarios we encountered in our experiments.


These candidates can suffer from a variety of errors. Most importantly, as explained in Section~\ref{sec:introduction}, we do not know if a given candidate conforms to the original natural-language assertion. Second, although it is true in this case, our solution approach also does not presume that the given assertion was valid for the method, and therefore we cannot simply filter for the valid formalizations (say, via testing).


\subsection*{Choosing Likely and Faithful Formalizations}

The second stage of our approach ranks and filters candidate formalizations using two complementary signals. First, we use a test generator to compute a \textbf{validity} score that captures whether an assertion is well-formed, and whether it is valid for the given method with respect to a set of tests generated by a test generator. Second, we compute a \textbf{conformance} score that estimates how well a candidate formalization covers the intent of the original NL specification.

The \textbf{validity} score is computed through a sequence of checks:
\begin{itemize}
    \item \textit{Syntactic check}: checks whether the candidate is syntactically and semantically well-formed, i.e., whether the formal assertion can compile without error 
    \item \textit{Fuzz-Safety check}: checks (using a fuzzer) whether the assertion is itself safe, i.e., that evaluating the assertion does not raise a runtime exception.
    \item \textit{Fuzz-Semantic check}: checks (again, using a fuzzer) whether the candidate is test-valid or test-invalid for the given method.
\end{itemize}

Note that our candidate assertions can use the \texttt{\textbackslash old(e)} operator to denote the value of the expression $e$ in the pre-state of the method. These are computed simply by memoizing the corresponding values in the pre-state.

In the context of our example, the validity checks eliminate Candidate \#4 in Table~\ref{tab:arraylist-add-candidates} because it has an extra parenthesis, making it syntactically invalid. Additionally, note that Candidate \#1 fails the fuzz-safety check since it does not use a null-safe interpretation of equality. 


The \textbf{conformance score} provides a complementary signal. To determine this score, we ask an LLM to describe the given formal assertion in natural language. We then ask an LLM to compare the clausal structure of this description against the original NL assertion and provide a measure of conformance between them (with $1$ being the best possible score). At a high level, candidates whose semantic content overlaps better with the original NL assertion receive higher scores, whereas those with missing, irrelevant, or conflicting content receive lower scores. This mechanism of using LLMs to provide numerical scores is known as LLM-as-a-Judge in the machine learning literature~\cite{zheng2023judgingllmasajudgemtbenchchatbot}. 

In our example, the LLM provides the following NL description for Candidate~\#1:

\begin{quote}
    \emph{``If the index is at least 0 and at most the size of the list before the insertion, then after the add operation the element at that index equals the inserted element.''}
\end{quote} 
This sentence is compared with the original specification for bidirectional clausal coverage, i.e., whether each substantive clause in one is represented in the other. In this case, the LLM judge assigns a perfect score of 1.0, because it does not find any differences. 

In contrast, the description for Candidate~\#3 begins as follows: \emph{``If the index is greater than or equal to 0 and less than the list size before the insertion...''}. The LLM judge then correctly identifies that ``less than the size'' misses an edge case compared to the original NL assertion, and therefore provides a lower score. 

We repeat the scoring process several times and average the LLM judge's scores. We set a threshold for the average conformance score and eliminate the candidates that do not meet this threshold. In our experiments we set the threshold to $0.6$, which  eliminates Candidate~\#3 in this example.

\begin{table}[t]
\centering
\footnotesize
\setlength{\tabcolsep}{4pt}
\renewcommand{\arraystretch}{1.12}
\begin{tabular}{lccccc}
\toprule
 & \textbf{Cand.~\#0} & \textbf{Cand.~\#1} & \textbf{Cand.~\#2} & \textbf{Cand.~\#3} & \textbf{Cand.~\#4} \\
\midrule
Syntactic check        & \checkmark & \checkmark & \checkmark & \checkmark & \xmark \\
Fuzz-Safety check     & \checkmark & \xmark & \checkmark & \checkmark & - \\
Fuzz-Semantic check        & \checkmark & -     & \xmark     & \xmark     & - \\
Validity score    & 1          & 0          & 0          & 0          & 0 \\
Conformance score & 1.000      & 1.000      & 0.875      & 0.525      & 1.000 \\
\bottomrule
\end{tabular}
\caption{Validity and conformance results for the example.}
\label{tab:arraylist-add-results}
\end{table}

Table~\ref{tab:arraylist-add-results} summarizes the validity and conformance results for the candidates. 

One of the key challenges in autoformalization is the inherent ambiguity in natural language. The ambiguity arises from several sources, among which is the fact that we do not know whether the original NL assertion is meant to be valid for the method. For example, when writing formal documentation, one may wish to always choose the formalization that is valid, whereas when attempting to find bugs in human or AI-written code, the more appropriate choice could be the formalization that reveals an assertion failure.

Therefore, among the remaining candidates we choose the candidate with the highest conformance score among those that passed the fuzz checks, and similarly one from those that failed the fuzz checks. In this case, that decision returns Candidate~\#0 and Candidate~\#1 as the most likely and faithful formalizations.



\subsection*{Final Disambiguation using Active Learning}

The final stage of our pipeline disambiguates between the likely and faithful autoformalizations using active learning. The idea is that the programmer can look at the distinction between the two candidates and determine the better formalization of the intended specification. In our work we utilize the modality of active learning where we produce a concrete valuation that distinguishes the two candidate formalizations. The user (or an automatic oracle) then determines whether the intended specification would satisfy this valuation, and we pick the corresponding formalization.

In this example, Candidate \#0 and Candidate \#1 are distinguished by the case where the inserted element is \texttt{null}. We use the test generator to come up with the distinguishing valuation. In our experiments, we simulate the user using an automatic oracle that is given the ground-truth, and answers using this ground truth. In this example, our framework ends up selects Candidate~\#0.

\section{Preliminaries}
\label{sec:prelim}

In this work we study the autoformalization of specifications in the realm of object-oriented programming. We now provide some background to formalize the vocabulary we employ in the paper. 

\paragraph*{Specification Language} Our specifications are written at the method level for methods in a class. 
We identify for each class a set of \emph{observer} methods, which are essentially side-effect-free methods (mathematical functions) that compute useful attributes of an object at a given state. For example, in our illustrative example in Section~\ref{sec:example} we utilized the observer method \texttt{size} in the \texttt{ArrayList} class. Observer functions are a standard feature in specification languages for object-oriented programs, especially in paradigms inspired from Design-by-Contract~\cite{dbc}. Observer functions are the only methods in a class that can be utilized in specifications. Specifications can also, of course, express properties of input parameters to a method and their return value.

At a high level, our assertion language is essentially the same as the Java Modeling Language (JML)~\cite{Leavens1999JMLAN}. Given a function with input variables $\overline{in}$ and output variables $\overline{out}$ (represented as $\mathit{\backslash result}$), our assertions can use the input and output variables, observer methods, implications, equality, Boolean operations, and certain basic operations for the various datatypes involved, e.g., inequalities and arithmetic operations for integer-typed values or specialized \texttt{.equals} functions for object types. We denote the values of observer functions $f(\mathit{args})$ on object variables $o$ by $o.f(\mathit{args})$. The special variable $\mathit{this}$ denotes the default current object for the class under consideration.

We also allow bounded quantification in assertions, e.g., $\forall i$ ranging over $0 \leq i < o.\mathit{size}()$ for an array $o$. Finally, we use the operator $\mathit{\backslash old}(e)$, which takes an expression $e$ and evaluates it in the pre-state of the method. For example, to say that the size of the object increases by $1$ compared to the pre-state, we would write $\mathit{this}.\mathit{size}() == \mathit{\backslash old}(\mathit{this}.\mathit{size}()) + 1$. 

\paragraph*{Test Generators} In this work we use test generators as verification oracles to judge the validity of assertions on methods. The test generator constructs a collection of well-formed inputs and checks the method under test for assertion violations within a given resource budget. In particular, it only constructs valid and reachable program states, by using constructors and other factory methods to construct valid objects and then calls the method under test. In our experiments we use Randoop~\cite{randoop}.

%




\section{Problem Statement and Methodology}
\label{sec:architecture}

\subsection*{Problem Statement}

We begin by stating our problem of study. We fix a logic $\mathcal{L}$ over which our framework is parameterized. In this work, the logic $\Ll$ is always ``executable'', that is, given a method and a formula in $\Ll$, it is possible to evaluate whether a given behavior of the method (inputs and outputs) satisfies the formula. Examples of such executable logics include the assertion libraries or sub-languages supported by any number of programming languages. In our experiments, we use an executable sublanguage of JML~\cite{Leavens1999JMLAN} as our logic for formalizing specifications.


Let us fix a module $M$ (we use class and module interchangeably) consisting of several methods $m$, each over a set of input variables $\overline{in}$ and output variables $\overline{out}$. The module may also optionally contain documentation strings $d_m$ for each method $m$. As explained in Section~\ref{sec:prelim}, in this work we consider specifications in the setting of object-oriented programs, so we also fix a formal object variable $o$ of the class $M$ to denote the calling object. 

\begin{definition}[Faithful Autoformalization of Specifications]
\label{def:problem}
Given a module $M$ and a natural-language specification $s$ for a method $m(\overline{in},\overline{out})$ in $M$, synthesize a formal specification $\psi(o,\overline{in},\overline{out})$ in $\Ll$ such that $\psi$ is a faithful formal translation of $s$.
\end{definition}

Note that techniques for solving the above problem are of course allowed to use all the information in $M$, including the signatures, documentation, and code for any of the methods. The above definition captures the minimal set of entities needed to define the problem.

Our definition above contains a couple of curiosities. First, we use the term \emph{faithful} to refer to the intuitive notion of a correct translation of a natural-language specification. Second, we do not ask for \emph{the} authoritative faithful formal translation, since natural language is inherently ambiguous. We discuss these challenges in detail below. We note here that in general the problem in Definition~\ref{def:problem} is under-specified, and we therefore utilize the standard practice of evaluating the performance of solutions to the problem on datasets containing ground-truth formal assertions.


\begin{figure*}[t] 
    \centering
    \includegraphics[width=0.8\textwidth]{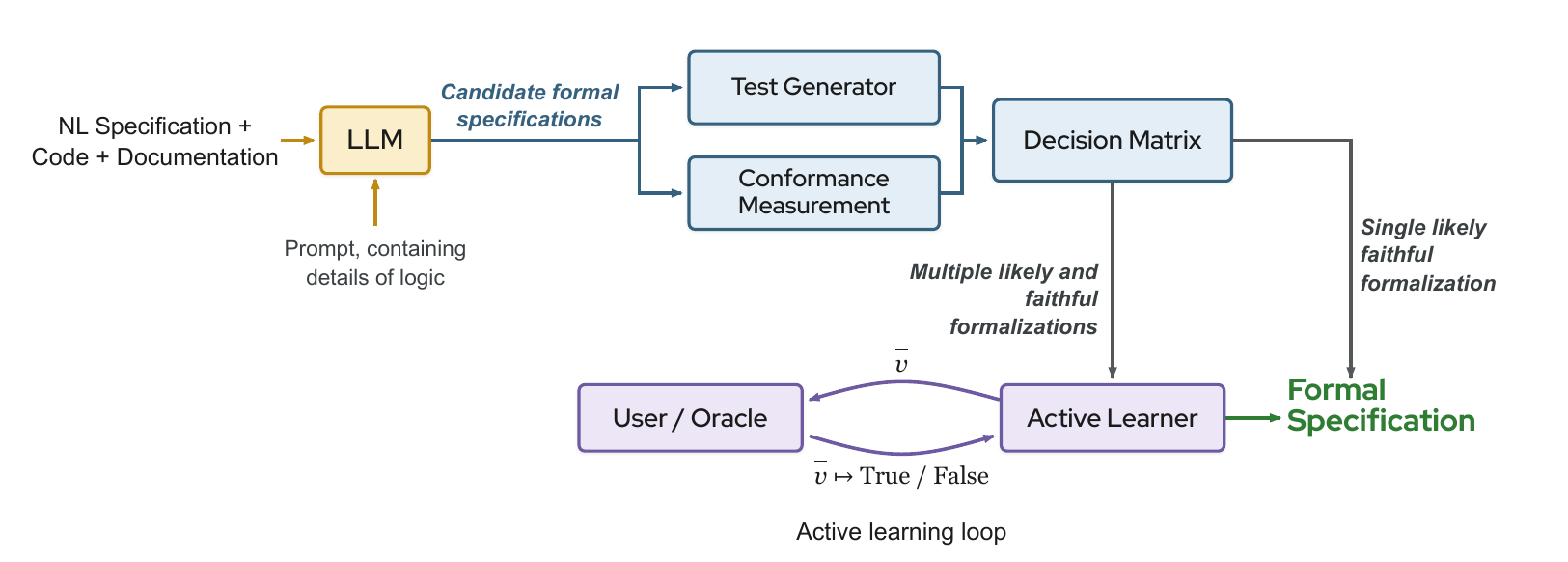} 
    \caption{The \tool architecture for faithful autoformalization of assertions}
    \label{fig:arch}
\end{figure*}

\subsection*{Challenges}

In this work we combine the powerful formalization abilities of LLMs and other tools such as fuzz testing and active learning to build techniques for autoformalizing natural-language assertions. We identify two key technical challenges in this process.

\mypara{Challenge 1: Identifying equivalence between natural language and formal language statements} The concept of ``equivalence'' between a natural language statement and a formula is intuitive to humans who are familiar with expressing formal logic. For example, $x > 0$ is a faithful translation of the statement ``$x$ is positive'', but $x <0$ is not. However, since the semantic content of an assertion can itself be complex and expressing formulas involves dealing with intricate logical operators, it can happen that LLMs do not faithfully translate natural-language assertions. They can suffer from a variety of issues, including specification language misuse, semantic overreach or overfitting to common cases, and depending on the model they may just plainly forget to formalize certain parts of a natural-language assertion~\cite{C2Stranslating}. 
One way of addressing this problem without performing post-training is to utilize a mechanizable definition of equivalence between a natural language utterance $s$ and a formula $\psi$ that approximates the intuitive equivalence as well as possible. One can then sample many outputs from an LLM, and filter out the candidates that do not pass the equivalence check. 
In this work we tackle the issue of identifying equivalence by introducing a technique called \emph{clausal coverage} conformance checking.

\mypara{Challenge 2: Tackling inherent ambiguity in natural language} The above challenge concerns the issue of tracking the semantic content that is explicitly present in an NL assertion. However, natural language is \emph{inherently} ambiguous, and it is not always possible to identify a single formal concept that corresponds to a natural language utterance. For example, people use the word ``positive'' in an informal sense to mean both strictly positive and non-negative, depending on the context. Another example documented in prior work on human understanding of LTL~\cite{shriram} is phrases like ``the red light is on until the blue light comes on'', which does not clearly specify whether the red light continues to be on or if it turns off when the blue light turns on. 
Note that both of these translations would be faithful to the original utterance, and are therefore not resolved by our solution to Challenge 1.



\subsection*{\tool Architecture}

We now describe \tool, our methodology for faithful autoformalization of specifications. Our overall pipeline is depicted in Figure~\ref{fig:arch}.

We first prompt an LLM with the module and method under consideration, as well as the NL specification $s$. We also provide information about the target logic in the prompt, and a few in-context examples. (see Appendix~\ref{app:prompt} for the full prompt.) 
We then obtain candidate formalizations $\psi_1, \psi_2,\ldots \psi_n$. The first stage of our architecture then calls, for each candidate formalization $\psi_i$, (a) a test generator that evaluates whether $\psi_i$ is test-valid for the given method, and (b) a conformance measurement module that measures the overlap in the semantic content between $\psi_i$ and the original NL assertion $s$. During this process we also eliminate syntactically malformed generations, or generations that do not compile without errors.

The second stage of our architecture is a decision matrix that takes the scores from the test generator and the conformance checker for all the candidates and returns the subset of them that are the most likely faithful formalizations. One of the key insights of our work is that inductive biases (e.g., presumption of validity) can be used to combat some of the inherent ambiguity in natural-language specifications. 

The final stage of our system is an active learning loop that disambiguates between the remaining candidates. We detail the construction of the nontrivial modules in our architecture below.

\smallskip
\mypara{Clausal Coverage Conformance Checker} One of the key contributions of our work is the construction of a conformance checker to address Challenge 1 above, namely the judgment of semantic equivalence between an NL assertion and a formal assertion. Since natural language is inherently ambiguous, we would not want the conformance checker to pick up on details in the formal assertion that were not specified clearly in the NL assertion. Our key insight here is that inherent ambiguity typically manifests \emph{locally} in the formal assertions, and it is therefore possible to rule out many bad formalizations by looking at the broad structure of the two assertions. 

We call the resulting technique \emph{clausal coverage}. We first ask an LLM to describe the given formalization in natural language. We then prompt the LLM to identify all the substantive clauses in the original NL assertion as well as the description, and ask whether the description is both (a) \textbf{Sound:} the description contains all the clauses in the NL assertion and (b) \textbf{Complete:} the description does not contain any information beyond what is given by the clauses in the NL assertion. We ask the LLM to provide a score in the interval $[-1,1]$ along with the following rubric: Positive scores indicate that the clauses in the original assertion were covered, with $+1$ indicating perfect conformance. Negative scores indicate that parts of the formalization conflict with the original assertion, with greater negative scores indicating more serious contradictions between the two assertions. A zero score is a neutral judgment that encompasses many cases like unrelated content in the formalization, or a failure to clearly detect either conformance or contradiction. This method of using an LLM as a scoring function is known as LLM-as-a-Judge in the Machine Learning literature~\cite{zheng2023judgingllmasajudgemtbenchchatbot}. We repeat this evaluation multiple times and average the scores, accepting a formalization as faithful only if it exceeds a certain threshold. In our experiments we repeat the check $8$ times and set the threshold to be $0.6$. We provide the prompt that we use for conformance checking in Appendix~\ref{app:prompt}.

Though the above technique is intuitive, to our knowledge we are the first to utilize this approach for conformance checking. In fact, in our initial version of the architecture we used a conformance checking method based on Natural Language Inference (NLI)~\cite{bowman2015largeannotatedcorpuslearning}, similar to the recommendations found in contemporary works that mention conformance checking~\cite{fazelnia2024lessonsusenaturallanguage}. These contemporary techniques also recommend doing a ``reverse translation'', and ask whether it is equivalent to the original input. However, we found that this only amplified the inherent ambiguity problem, and LLMs turned out to be overly sensitive to differences in phrasing. We evaluate ablations of our conformance checking technique in Section~\ref{sec:eval}.

\mypara{Decision Matrix and Inductive Biases} The second insight of our work is that the inherent ambiguity in natural language can be addressed in part by incorporating inductive biases specific to the application domain. For example, if the statement ``$x$ is positive'' yields both $x > 0$ and $x \geq 0$ as faithful candidate formalizations where only one of them is test-valid for the given method, we may be able to pick the valid one if we assume that the application demands a presumption of validity of the assertion on the part of the user writing the NL specification. This is often the case in documentation of code, and has been the default incorporated by other related works on specification mining (see Section~\ref{sec:relwork}). In contrast, if the user is writing assertions to identify edge cases and finding bugs, we may in fact want to choose the formalization that is \emph{not} test-valid for the given method, and return the inputs that raise the assertion violation.

In our work we assume a more general setting with no presumptions about the validity of the assertions. Instead, we separate the generated assertions into test-valid and test-invalid ones, and choose the candidates with the highest conformance score from each category. Of course, we do not pick any candidates whose conformance score is below the acceptable threshold. The intuition behind our approach is that the best test-valid and test-invalid formalizations together capture the range of ambiguity in the original natural-language assertion. This is similar to the idea of a version space bounded by a general hypothesis and a specific hypothesis in traditional work on learning theory~\cite{MITCHELL1977VERSIONSPACES}.

\mypara{Active Learning Loop} The final stage of our pipeline utilizes active learning to disambiguate between faithful candidate formalizations. The intuition is that the remaining candidates at this stage are the likely and faithful formalizations, and indicate the range of inherent ambiguity or under-specification in the original NL assertion. In this work we use the active learning paradigm that utilizes a teacher (typically the user) who can say whether a given data point satisfies the intended assertion. A data point here is a valuation over the entities/variables involved in the assertion. So, for example, if the original assertion is ``If the inserted element is positive then the size of the array increases by 1'', the valuation would provide values for the inserted element, as well as pre- and post-state values for the size of the array object. 

In this work the active learning only distinguishes between the two candidates that are returned by the decision matrix. We compare them using the test generator to produce a valuation ($\overline{v}$ in Figure~\ref{fig:arch}) that distinguishes them. In Section~\ref{sec:example} this valuation was setting the inserted element to \texttt{null}. In particular, we ensure that the valuation is \emph{realizable}, i.e., one that can actually be produced by the method's transformation. For example, if the valuation involves both the old and new sizes of an object, we only ask the test generator to search over values for the old size, and execute the program to yield the size in the post-state. We then query the user to gather whether the intended formal assertion satisfies the distinguishing point, and pick the appropriate candidate. In our experiments we utilize datasets with ground-truth formal assertions and simulate the user by executing the ground-truth assertion on the distinguishing valuation. 

Of course, it can happen that the decision matrix only produces one candidate that exceeds the conformance threshold, and we simply return that candidate in this case. One interesting scenario is when there are more than two candidates with sufficiently high conformance scores. One can then consider distinguishing between these using active learning. We do not entertain this scenario in this work, but it would be simple to make a sufficient number of pairwise comparisons (either all pairs or perhaps sequentially) and find distinguishing inputs as we currently do. We can also consider minimizing the number of distinguishing points to reduce user involvement, and there has been some recent work on this problem~\cite{Dillig-PLDI26ActiveLearning}.

\section{Benchmarking}
\label{sec:impl}


We curate a suite of benchmarks for evaluating our approach from existing datasets. In particular, as we discuss in Section~\ref{sec:relwork}, existing benchmarks are not precisely aligned with our task of formalizing assertions, and hence we need some adaptation/curation.
As such, the datasets we develop are one of the contributions of this work.

Our datasets are all centered around the formal assertions in the prior work of Zhai et al. called C2S~\cite{C2Stranslating}. The dataset published in this work contained a set of Java collection-style classes and formal specifications for the methods in those classes. We adapted this dataset, manually resolved errors in some of the formal assertions, and augmented it with more collection-style classes and formal specifications.

\noindent\textbf{C2S-Augmented Synthetic Dataset.} The first benchmark suite we develop is one where the NL specifications are synthetically generated from the formal assertions in the augmented C2S suite we describe above. We prompt Claude-Sonnet-4.6 to describe the formal specifications with concise and natural-sounding English statements. We obtain a suite of 416 (NL specification, ground-truth formal specification) pairs across 19 classes as a result. 

\noindent\textbf{Buggy Dataset.} The specifications in the augmented C2S suite are valid assertions. We therefore construct a Buggy assertion dataset to evaluate translation of natural-language specifications when specifications are incorrect or program contexts are inconsistent. This dataset is derived from a subset of the 19 classes described above. We create two variants: a buggy-code variant where we perturbed the code for the target method, and a buggy-assertion variant where we perturbed the specifications into incorrect or misleading assertions. Once again, we use Claude Sonnet 4.6 to produce the NL specifications for the buggy assertions.

\noindent\textbf{Manual NL Specs Dataset.} We also create a smaller dataset where we write the natural-language specifications manually. We choose three additional Java classes not included in the above set along with formal specifications. We then asked one of the authors to write specifications, with the only instruction being that they could look at the code and the documentation and needed to write an NL specification that would essentially capture the given formal specification. This author was not involved in the identification of the three additional classes or the formal specifications, and was also not shown any of the synthetically generated NL specifications from the above datasets. The author was, however, aware of the solution approach in the paper. This dataset contains 39 (NL, formal) specification pairs.

We provide a detailed account of the dataset statistics in Appendix~\ref{app:dataset}.


\section{Evaluation}
\label{sec:eval}

\begin{table*}[t]
\centering
\footnotesize
\setlength{\tabcolsep}{3pt}
\renewcommand{\arraystretch}{1.12}
\begin{tabular}{llccccc}
\textbf{Dataset} & \textbf{Model} & \textbf{Any-Equiv} & \textbf{Oneshot} & \textbf{\tool-Acc} & \textbf{\tool-Prec} & \textbf{\tool-Rec} \\
\midrule
\multirow{2}{*}{C2S-Aug. (416)} 
& GPT-oss & 91.8\% (382/416) & 85.3\% (355/416) & 89.7\% (373/416) & 92.8\% (373/402) & 97.6\% (373/382) \\
& Qwen2.5-Coder & 81.3\% (338/416) & 75.0\% (312/416) & 75.7\% (315/416) & 91.6\% (315/344) & 93.2\% (315/338) \\
\midrule
\multirow{2}{*}{Buggy Code (20)} 
& GPT-oss & 75.0\% (15/20) & 60.0\% (12/20) & 70.0\% (14/20) & 70.0\% (14/20) & 93.3\% (14/15) \\
& Qwen2.5-Coder & 65.0\% (13/20) & 60.0\% (12/20) & 60.0\% (12/20) & 75.0\% (12/16) & 92.3\% (12/13) \\
\midrule
\multirow{2}{*}{Buggy Asrt. (66)} 
& GPT-oss & 69.7\% (46/66) & 54.5\% (36/66) & 57.6\% (38/66) & 61.3\% (38/62) & 82.6\% (38/46) \\
& Qwen2.5-Coder & 28.8\% (19/66) & 16.7\% (11/66) & 7.6\% (5/66) & 22.7\% (5/22) & 26.3\% (5/19) \\
\midrule
\multirow{2}{*}{Manual NL (39)}
& GPT-oss & 87.2\% (34/39) & 74.4\% (29/39) & 84.6\% (33/39) & 86.8\% (33/38) & 97.1\% (33/34) \\
& Qwen2.5-Coder & 69.2\% (27/39) & 64.1\% (25/39) & 59.0\% (23/39) & 85.2\% (23/27) & 85.2\% (23/27) \\
\bottomrule
\end{tabular}
\caption{Effect of the decision matrix on selecting candidates with GPT-oss and Qwen2.5-Coder backbone models across datasets. Counts are shown in parentheses.}
\label{tab:main-res}
\end{table*}

We follow a routine experimental setup described in Appendix~\ref{app:experimental-setup} and evaluate our implementation of \tool on the following research questions:
\begin{itemize}
    \item \textbf{RQ1:} \emph{How effective is our approach at autoformalizing natural-language specifications?} 

    \item \textbf{RQ2:} \emph{What, if any, are the improvements our approach provides over purely calling an LLM to solve the task?} 
    
    \item \textbf{RQ3:} \emph{What is the effectiveness of our approach on manually written NL specifications?} 
    
    \item \textbf{RQ4:} \emph{How does the choice of backbone LLM affect autoformalization performance?} 
    
    \item \textbf{RQ5:} \emph{How effective is conformance checking in improving the faithfulness of formalizations? How does our approach compare to baselines?} 

    \item \textbf{RQ6:} \emph{How effective is active learning in the architecture?} 
\end{itemize}

We use the following task-level metrics throughout the evaluation (see e.g., Table~\ref{tab:main-res}).
\textbf{Any-Equiv} measures the generation potential of the backbone LLM: it reports whether the candidate pool contains at least one assertion equivalent to the ground truth. This is essentially the Pass@5 metric.
\textbf{Oneshot} is the Pass@1 metric: it measures whether the first generated candidate from the LLM is equivalent to the ground-truth assertion. This provides the baseline comparison of plainly calling an LLM to solve the task at hand.
\textbf{\tool-Acc} measures the end-to-end accuracy of our approach: the final assertion selected by the full pipeline (and hence passing the conformance threshold 0.6) is logically equivalent to the ground truth.
\textbf{\tool-Prec} measures the reliability of \tool outputs: among all tasks for which \tool returns a high-conformance ($\ge0.6$) final assertion, it reports the fraction whose final assertion is equivalent. This is a \emph{precision} metric for the tool. In other words, when \tool returns a formal assertion, how likely is it to be correct? This is one of the most important metrics as it measures how accurate the tool is when it reports a formalization. Finally, \textbf{\tool-Rec} measures recall, i.e., how often \tool recovers a correct assertion when one is available: among tasks whose candidate pool contains at least one equivalent assertion, it reports the fraction for which \tool selects a logically equivalent high-conformance (score$\ge0.6$) final assertion. 

We note here that Section~\ref{sec:relwork} discusses several works on the related problem of mining formal specifications from natural language documentation using LLMs. To our knowledge, all these works either (a) plainly use an LLM, or (b) use an LLM to generate formal assertions and then weed out invalid ones using a fixed set of tests/test generator and a presumption of correctness. Our experiments essentially capture (a) using the Oneshot metric, and (b) via our evaluation ablating the conformance check (but only on benchmarks that have valid assertions). Another technique in the literature is the use of an LLM to `repair' test-invalid assertions. We do not adopt this since we do not have a presumption of validity. 

\subsection*{RQ1: How effective is our approach at autoformalizing natural-language specifications?}

Table~\ref{tab:main-res} summarizes the overall effectiveness of \tool across datasets and open-source backbone models. The results show two complementary effects. First, the LLMs are often able to generate at least one suitable candidate formalization, as reflected by the Any-Equiv metric. Second, given the generated candidate pool, our decision matrix and active-learning resolver effectively select a final candidate with high precision and recall. On the main C2S-Augmented dataset, GPT-oss achieves an overall success rate of 89.7\%, while Qwen2.5-Coder achieves 75.7\%.

The same trend holds across the additional datasets. On Buggy Code, \tool maintains high recall for both models. Performance is lower on Buggy Assertion, especially for Qwen2.5-Coder, reflecting the difficulty of formalizing misleading or incorrect specifications. This also suggests that LLMs have a noticeable bias toward producing valid-looking assertions or specifications, making it harder for them to synthesize intentionally invalid assertions from natural-language descriptions if one were to use them for, say, bug finding. Overall, these results indicate that \tool is effective across the evaluated datasets.

We also evaluated our pipeline on a variant of the above dataset where the NL specifications were generated by GPT-5, and we found that the results were materially no different from the ones presented in Table~\ref{tab:main-res}.

\subsection*{RQ2: What, if any, are the improvements our approach provides over purely calling an LLM to solve the task?}

This is the main research question of this paper. With any methodologies that augment LLMs, the main baseline for comparison is the one that a naive user would resort to, namely simply calling an LLM with the task description and choosing the first response. While we do not use a naive prompt, we do capture this approach using the Oneshot metric. 

Table~\ref{tab:main-res} shows that our approach ({\bfseries \tool-Acc}) generally performs improves over the \textbf{Oneshot} metric. On the main C2S-Augmented dataset, GPT-oss improves from an \textbf{Oneshot} accuracy of 85.3\% to a {\bfseries \tool-Acc} accuracy of 89.7\%, while Qwen2.5-Coder improves slightly from 75.0\% to 75.7\%. The improvement is consistent on other datasets, especially for Buggy Code, where GPT-oss increases from 60.0\% to 70.0\%.

The more interesting results lie in the {\bfseries \tool-Prec} scores. Note that since LLMs always provide a response, the \textbf{Oneshot} metric is also the same as the precision metric for plainly using an LLM. Table~\ref{tab:main-res} shows that our approach significantly improves precision. The effect is more pronounced for smaller models: for Qwen2.5-Coder the precision improves from 75\% to 91.6\% on C2S-Aug, and from 64.1\% to 85.2\% on the manual specs dataset. Even on GPT-oss, which is a larger model, the precision increases from 80\% to 88\% on average. This is a significant difference and a promising observation, since practical adoption typically requires more than 90\% precision (and more than 95\% in many cases). 

Table~\ref{tab:main-res} also shows that the precision is improved without hurting recall. On C2S-Augmented, while the precision reaches 92.8\% for GPT-oss and 91.6\% for Qwen2.5-Coder, recall reaches 97.6\% and 93.2\%, respectively. Recall captures the complementary property to precision: when the candidate pool contains a correct formalization, a high recall indicates that the filtering process usually preserves it rather than discarding it. Together, these results suggest that the decision matrix improves reliability without substantially reducing the availability of useful outputs.


\subsection*{RQ3: What is the effectiveness of our approach on manually written NL specifications?}

Machine-generated natural-language specifications can be more regular and templated compared to human-written specifications written by real developers. This raises the concern that evaluation on such data may overestimate performance if autoformalizing models exploit machine-generated phrasing patterns. To address this concern, we evaluate \tool on the Manual NL Specs dataset, where the natural-language specifications are written manually. As shown in Table~\ref{tab:main-res}, our tool still achieves meaningful end-to-end performance on these specifications: GPT-oss obtains an accuracy of 84.6\%, while Qwen2.5-Coder obtains 59.0\%. This suggests that the performance of \tool may be generalizable to real-world use cases with human users.

The precision and recall results further clarify how \tool behaves in this setting. For GPT-oss, the reported outputs have a precision of 86.8\%, and the tool also recovers 97.1\% of cases where an equivalent candidate is available. Qwen2.5-Coder is weaker on this dataset, but its reported outputs still have relatively high precision, 85.2\%, with a recall of 85.2\%.

\subsection*{RQ4: How does the choice of backbone LLM affect autoformalization performance?}

\begin{table}[t]
\centering
\footnotesize
\setlength{\tabcolsep}{10pt}
\renewcommand{\arraystretch}{1.2}
\begin{tabular}{lccc}
\textbf{Model} & \textbf{Any-Equiv} & \textbf{Oneshot} & \textbf{\tool-Prec} \\
\midrule
Qwen2.5-Coder & 81.2\% (78/96) & 75.0\% (72/96) & 88.9\% (72/81) \\
Qwen3-235B & 90.6\% (87/96) & 75.0\% (72/96) & 89.0\% (81/91) \\
GPT-oss & \textbf{93.8\% (90/96)} & 77.1\% (74/96) & 93.5\% (87/93) \\
GPT-5.5 & 91.7\% (88/96) & 88.5\% (85/96) & \textbf{95.6\% (87/91)} \\
Claude-Opus-4.8 & 92.7\% (89/96) & \textbf{90.6\% (87/96)} & 92.6\% (88/95) \\
\bottomrule
\end{tabular}
\caption{RQ4: Backbone model comparison on a subset of the C2S-Aug. dataset. Counts are shown in parentheses.}
\label{tab:rq4-model-comparison}
\end{table}

Table~\ref{tab:rq4-model-comparison} compares different backbone models on a randomly chosen subset of the C2S-Augmented dataset. We choose a subset owing to resource constraints. In addition to the two main open-source models, we evaluate three stronger backbones: \textsc{Qwen3-235B-A22B-Instruct-2507}~\cite{qwen3technicalreport}, \textsc{GPT-5.5}~\cite{openai2026gpt55}, and \textsc{Claude-Opus-4.8}~\cite{anthropic2026claudeopus48}. Qwen3-235B is a larger model from the Qwen family, while GPT-5.5 and Claude-Opus-4.8 represent closed-source frontier models. For GPT-5.5, we use the medium effort.

The results show that backbone quality has a clear effect on candidate generation. The smaller Qwen2.5-Coder model obtains an Any-Equiv score of 81.2\%, whereas the larger or stronger models all exceed 90\%, indicating that they are more likely to generate at least one equivalent formalization in the candidate pool. The Oneshot results vary even more substantially: Claude-Opus-4.8 and GPT-5.5 achieve the strongest first-candidate accuracy, 90.6\% and 88.5\%, respectively, while GPT-oss and the Qwen models are lower. At the same time, the precision remains high across all models, suggesting that the decision matrix can still select reliable outputs even when the first generated candidate is not consistently correct. Overall, stronger backbones improve raw generation quality, while \tool improves the reliability of the final reported formalizations across model families.

\subsection*{RQ5: How effective is conformance checking in improving the faithfulness of formalizations? How does our approach compare to baselines?}

\begin{table}[t]
\centering
\footnotesize
\setlength{\tabcolsep}{10pt}
\renewcommand{\arraystretch}{1.05}
\begin{tabular}{llccc|cc}
\textbf{Dataset} & \textbf{Model} 
& \multicolumn{3}{c|}{\textbf{Conf. Classifier}} 
& \multicolumn{2}{c}{\textbf{Active Learning}} \\
\cmidrule(lr){3-5}\cmidrule(lr){6-7}
& & \textbf{Prec} & \textbf{Rec} & \textbf{F1} 
& \textbf{Cases} & \textbf{Acc} \\
\midrule
\multirow{2}{*}{C2S-Aug.}
& GPT-oss & 88.1\% & 99.2\% & 93.3\% & 9 & 100.0\% \\
& Qwen2.5 & 79.9\% & 91.8\% & 85.4\% & 9 & 77.8\% \\
\midrule
\multirow{2}{*}{Buggy Code}
& GPT-oss & 58.6\% & 100.0\% & 73.9\% & 0 & -- \\
& Qwen2.5 & 61.8\% & 91.7\% & 73.8\% & 0 & -- \\
\midrule
\multirow{2}{*}{Buggy Asrt.}
& GPT-oss & 61.6\% & 97.6\% & 75.5\% & 5 & 60.0\% \\
& Qwen2.5 & 18.6\% & 32.1\% & 23.5\% & 6 & 83.3\% \\
\midrule
\multirow{2}{*}{Manual NL}
& GPT-oss & 80.2\% & 91.0\% & 85.3\% & 1 & 100.0\% \\
& Qwen2.5 & 64.1\% & 83.3\% & 72.5\% & 1 & 100.0\% \\
\bottomrule
\end{tabular}
\caption{Conformance-threshold classifier performance and active-learning accuracy across main datasets. The conformance classifier is evaluated on all generated candidates with threshold 0.6. Active-learning accuracy is evaluated only on eligible ambiguous cases.}
\label{tab:conf-active}
\end{table}

\begin{figure}[t]
    \centering
    \includegraphics[width=0.4\linewidth]{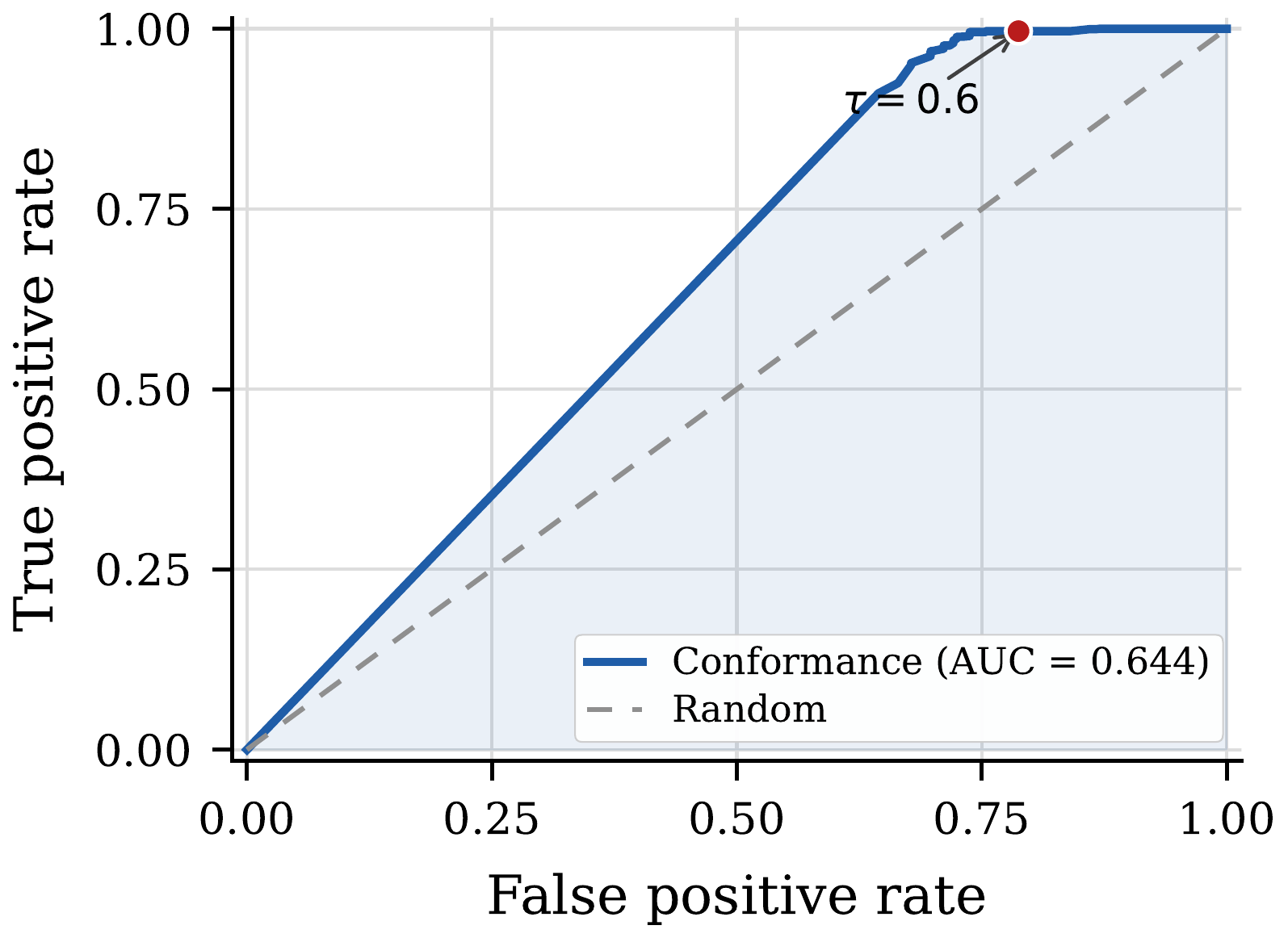}
    \caption{ROC curve for conformance-threshold classifier on all of the generated candidates in C2S-Aug. dataset with GPT-oss backbone model.}
    \label{fig:roc-c2saug-gptoss}
\end{figure}

To isolate the effect of conformance checking, we evaluate it as a classifier: a candidate is predicted correct when its conformance score is at least \(\tau\), and the label is logically equivalent to the ground-truth specification. Figure~\ref{fig:roc-c2saug-gptoss} shows the ROC curve obtained by sweeping the conformance threshold \(\tau\); based on this tradeoff, we set \(\tau=0.6\), because it favors high recall while still filtering low-conformance candidates.

Table~\ref{tab:conf-active} reports performance at this threshold. On C2S-Augmented, conformance achieves strong F1 scores for GPT-oss and Qwen2.5-Coder, 93.3\% and 85.4\%, respectively. Recall is consistently high, showing that the filter rarely removes correct candidates. Precision is lower, confirming that conformance is not a correctness oracle, but still provides a useful signal for ranking and filtering. Qwen2.5-Coder performs poorly on Buggy Assertion, suggesting that smaller models struggle when many candidates look valid but do not match the intended specification.

Table~\ref{tab:conf-impl-auc} compares conformance implementations using AUC. Detailed explanation and visualization can be found in Appendix~\ref{app:conf-impl}. Our full method achieves the best aggregate AUC, 0.659, outperforming other methods. The gap between Ours and Ours-noCtx shows that contextual information improves ranking quality. Although NLI performs well on Manual NL, its overall AUC and fixed-threshold behavior are weaker. These results support the use of a contextual, roundtrip-based bidirectional clausal conformance check.

\begin{table}[t]
\centering
\footnotesize
\setlength{\tabcolsep}{5pt}
\renewcommand{\arraystretch}{1.2}
\begin{tabular}{lcccccccc}
\textbf{Method} & \textbf{RT} & \textbf{Bi} & \textbf{Ctx} & \textbf{C2S-S} & \textbf{Bug-C} & \textbf{Bug-A} & \textbf{Manual} & \textbf{All} \\
\midrule
Direct & \xmark & \xmark & \cmark & 0.567 & 0.571 & 0.690 & 0.559 & 0.622 \\
Simple & \cmark & \xmark & \xmark & 0.565 & 0.464 & 0.682 & 0.596 & 0.610 \\
Bi-NLI & \cmark & \cmark & \xmark & 0.554 & 0.562 & 0.614 & \textbf{0.737} & 0.608 \\
Ours-noCtx & \cmark & \cmark & \xmark & 0.535 & \textbf{0.600} & 0.715 & 0.636 & 0.618 \\
Ours & \cmark & \cmark & \cmark & \textbf{0.611} & 0.510 & \textbf{0.744} & 0.664 & \textbf{0.659} \\
\bottomrule
\end{tabular}
\caption{AUC comparison of conformance methods on all generated candidates with GPT-oss as the backbone LLM model. RT denotes roundtrip, Bi denotes bidirectional, and Ctx denotes context.}
\label{tab:conf-impl-auc}
\end{table}

\subsection*{RQ6: How effective is active learning in the architecture?}

We evaluate the effectiveness of active learning across datasets. The learner is only invoked when the decision matrix identifies both a valid candidate and an invalid candidate. We find in our experiments that active learning only applies in a few cases in our datasets. In Table~\ref{tab:conf-active}, \textit{Cases} counts eligible active-learning tasks: tasks where the learner is invoked and at least one of the two competing candidates is equivalent to the ground truth. Accuracy measures whether the learner selects an equivalent candidate among these eligible cases. Although in practice it is useful to have active learning interactions be minimal to prevent user fatigue, we believe that a larger study may be needed to showcase a more complete picture of the prevalence of cases where active learning would be required. 

\section{Related Work}
\label{sec:relwork}

\subsection*{Traditional Specification Mining and Contract Inference}

Traditional specification mining and contract inference aim to recover program properties automatically. Early passive techniques infer invariants or API protocols from observed executions and traces, such as Daikon-style dynamic invariant detection and trace-based protocol mining~\cite{ernst1999dynamically,ernst2007daikon,ammons2002mining,whaley2002automatic,xie2006automatic,alur2005synthesis}. These methods are grounded in program behavior, but can only infer properties exposed by the observed executions and generally do not capture developer intent directly.

Later work combines dynamic analysis with testing, symbolic reasoning, predicate synthesis, and learning. For example, Sankaranarayanan et al.~\cite{sankaranarayanan2008dynamic} combine dynamic and symbolic techniques for invariant inference, Padhi et al.~\cite{padhi2016data} synthesize preconditions from execution data, and Astorga et al.~\cite{astorga2019learning,astorga2021synthesizing} learn stateful and object-oriented contracts with guarantees defined relative to a test generator. More recent work also mines specifications from natural-language API documentation and comments~\cite{pandita2012inferringmethodspec,C2Stranslating}. Our work is complementary: we start from a natural-language assertion and study whether LLMs can autoformalize it to an executable assertion while preserving intent. Unlike prior mining approaches, \tool explicitly evaluates generated candidates along two dimensions: their execution-level validity and their semantic alignment with the user's intended specification.

\subsection*{LLM-Based Specification Generation and Autoformalization}

LLM-based specification generation has emerged along two main directions. The first translates natural-language requirements, documentation, or informal descriptions into formal artifacts such as regular expressions, first-order logic, temporal logic, contracts, and hardware assertions~\cite{hahn2022formalspecificationsnaturallanguage,cosler2023nl2spec,yan2024assertllm,xia2026doc2spec,richter2025nl2contract}. These systems show that learned models can recover useful semantic structure from natural language, often with grammars, templates, interaction, or refinement to improve syntactic validity and reduce ambiguity.

The second direction generates specifications directly from code or program context. SpecGen~\cite{ma2025specgenautomatedgenerationformal} synthesizes formal specifications with LLMs and improves them through mutation-based refinement and verification-guided selection. SpecSyn~\cite{specsyn} targets real-world program verification by decomposing programs and refining generated specifications based on semantic strength. SLD-Spec~\cite{sldspec} focuses on complex loop functions, using program slicing and LLM-based logical deletion to remove irrelevant or incorrect specifications. AutoReSpec~\cite{autorespec} further explores validator-guided collaborative LLM generation for improving verifiable specifications. Recent work has also used LLM-based extraction of specifications for libraries to prove a client correct~\cite{uppar2026verificationmodulotestedlibrary}.

Our work differs in that it targets module-level code contexts and explicitly starts from natural-language specifications, with the goal of formalizing them into executable assertions while preserving the user's intended meaning.

\subsection*{Round-Trip Techniques for Specification and Verification}

Round-trip techniques translate artifacts back into the original representation to expose information loss or semantic mismatch. In NL-to-formal-specification settings, prior systems use back-translation or subformula-to-text mappings to inspect, debug, or refine generated specifications~\cite{cosler2023nl2spec}. Clover checks consistency among code, docstrings, and formal annotations through reconstruction tests~\cite{sun2024clover}, while Claimcheck back-translates verified Dafny lemmas to detect proof--intent mismatches~\cite{claimcheck}. Round-trip translation has been used for quality control in machine translation and software traceability, but is not always a reliable proxy for forward-translation correctness~\cite{somers2005roundtrip}.

Our conformance check adapts this idea to assertion autoformalization. Rather than simply prompting an LLM to judge equivalence, it evaluates bidirectional clause-level coverage with relevant code context, including method signatures, arguments, return values, and documentation.

\section{Conclusions}
\label{sec:conclusion}
We present \tool, a pipeline for autoformalizing natural-language specifications into executable assertions by combining LLM-based candidate generation with a decision matrix based on execution-level validity and semantic-level conformance, together with an active-learning loop for disambiguation. Rather than relying on a single raw LLM output, \tool generates multiple candidate formalizations, checks their executable behavior, estimates their alignment with the intended specification, and selects a reliable final formalization when possible. Across multiple datasets and backbone models, our results show that \tool improves the reliability of the formalizations it reports. The ablation studies further show that contextual, roundtrip-based bidirectional clausal coverage provides an effective conformance signal for candidate selection.


\bibliographystyle{plain}
\bibliography{references}

@inproceedings{C2Stranslating,
author = {Zhai, Juan and Shi, Yu and Pan, Minxue and Zhou, Guian and Liu, Yongxiang and Fang, Chunrong and Ma, Shiqing and Tan, Lin and Zhang, Xiangyu},
title = {C2S: translating natural language comments to formal program specifications},
year = {2020},
isbn = {9781450370431},
publisher = {Association for Computing Machinery},
address = {New York, NY, USA},
url = {https://doi.org/10.1145/3368089.3409716},
doi = {10.1145/3368089.3409716},
booktitle = {Proceedings of the 28th ACM Joint Meeting on European Software Engineering Conference and Symposium on the Foundations of Software Engineering},
pages = {25–37},
numpages = {13},
location = {Virtual Event, USA},
series = {ESEC/FSE 2020}
}

@article{dbc,
author={Meyer, Bertrand},
journal={ Computer },
title={{ Applying "Design by Contract" }},
year={1992},
volume={25},
number={10},
ISSN={1558-0814},
pages={40-51},
doi={10.1109/2.161279},
url = {https://doi.ieeecomputersociety.org/10.1109/2.161279},
publisher={IEEE Computer Society},
address={Los Alamitos, CA, USA},
month=oct
}

@misc{meyer1992eiffel,
  title={Eiffel: The Language. Object-Oriented Series},
  author={Meyer, Bertrand},
  year={1992},
  publisher={Prentice Hall, New York, NY}
}

@inproceedings{Leavens1999JMLAN,
  title={JML: A Notation for Detailed Design},
  author={Gary T. Leavens and Albert L. Baker and Clyde Ruby},
  booktitle={Behavioral Specifications of Businesses and Systems},
  year={1999},
  url={https://api.semanticscholar.org/CorpusID:10401486}
}

@inproceedings{specsharp,
author = {Barnett, Mike and Leino, K. Rustan M. and Schulte, Wolfram},
title = {The spec\# programming system: an overview},
year = {2004},
isbn = {3540242872},
publisher = {Springer-Verlag},
address = {Berlin, Heidelberg},
url = {https://doi.org/10.1007/978-3-540-30569-9_3},
doi = {10.1007/978-3-540-30569-9_3},
booktitle = {Proceedings of the 2004 International Conference on Construction and Analysis of Safe, Secure, and Interoperable Smart Devices},
pages = {49–69},
numpages = {21},
location = {Marseille, France},
series = {CASSIS'04}
}

@inproceedings{codecontracts,
author = {Logozzo, Francesco},
title = {Practical specification and verification with code contracts},
year = {2013},
isbn = {9781450324670},
publisher = {Association for Computing Machinery},
address = {New York, NY, USA},
url = {https://doi.org/10.1145/2527269.2534188},
doi = {10.1145/2527269.2534188},
booktitle = {Proceedings of the 2013 ACM SIGAda Annual Conference on High Integrity Language Technology},
pages = {7–8},
numpages = {2},
location = {Pittsburgh, Pennsylvania, USA},
series = {HILT '13}
}

@misc{Lahiri,
      title={Intent Formalization: A Grand Challenge for Reliable Coding in the Age of AI Agents}, 
      author={Shuvendu K. Lahiri},
      year={2026},
      eprint={2603.17150},
      archivePrefix={arXiv},
      primaryClass={cs.SE},
      url={https://arxiv.org/abs/2603.17150}, 
}

@misc{kiro,
  author       = {{Kiro}},
  title        = {Introducing Kiro},
  year         = {2025},
  howpublished = {\url{https://kiro.dev/blog/introducing-kiro/}},
  note         = {Accessed: 2026-06-30}
}

@misc{githubspeckit,
  author       = {{GitHub}},
  title        = {Spec-Driven Development with AI: Get Started with a New Open Source Toolkit},
  year         = {2025},
  howpublished = {\url{https://github.blog/ai-and-ml/generative-ai/spec-driven-development-with-ai-get-started-with-a-new-open-source-toolkit/}},
  note         = {Accessed: 2026-06-30}
}

@misc{hahn2022formalspecificationsnaturallanguage,
  title={Formal Specifications from Natural Language},
  author={Christopher Hahn and Frederik Schmitt and Julia J. Tillman and Niklas Metzger and Julian Siber and Bernd Finkbeiner},
  year = 2022,
  month = 6,
  doi = "10.48550/arxiv.2206.01962"
}

@inproceedings{ma2025specgenautomatedgenerationformal,
author = {Ma, Lezhi and Liu, Shangqing and Li, Yi and Xie, Xiaofei and Bu, Lei},
title = {SpecGen: Automated Generation of Formal Program Specifications via Large Language Models},
year = {2025},
isbn = {9798331505691},
publisher = {IEEE Press},
url = {https://doi.org/10.1109/ICSE55347.2025.00129},
doi = {10.1109/ICSE55347.2025.00129},
booktitle = {Proceedings of the IEEE/ACM 47th International Conference on Software Engineering},
pages = {16–28},
numpages = {13},
location = {Ottawa, Ontario, Canada},
series = {ICSE '25}
}

@inproceedings{wen2024autospec,
author = {Wen, Cheng and Cao, Jialun and Su, Jie and Xu, Zhiwu and Qin, Shengchao and He, Mengda and Li, Haokun and Cheung, Shing-Chi and Tian, Cong},
title = {Enchanting Program Specification Synthesis by Large Language Models Using Static Analysis and Program Verification},
year = {2024},
isbn = {978-3-031-65629-3},
publisher = {Springer-Verlag},
address = {Berlin, Heidelberg},
url = {https://doi.org/10.1007/978-3-031-65630-9_16},
doi = {10.1007/978-3-031-65630-9_16},
booktitle = {Computer Aided Verification: 36th International Conference, CAV 2024, Montreal, QC, Canada, July 24–27, 2024, Proceedings, Part II},
pages = {302–328},
numpages = {27},
location = {Montreal, QC, Canada}
}

@inproceedings{zhong2009inferringresourcespec,
author = {Zhong, Hao and Zhang, Lu and Xie, Tao and Mei, Hong},
title = {Inferring Resource Specifications from Natural Language API Documentation},
year = {2009},
isbn = {9780769538914},
publisher = {IEEE Computer Society},
address = {USA},
url = {https://doi.org/10.1109/ASE.2009.94},
doi = {10.1109/ASE.2009.94},
booktitle = {Proceedings of the 24th IEEE/ACM International Conference on Automated Software Engineering},
pages = {307–318},
numpages = {12},
series = {ASE '09}
}

@inproceedings{pandita2012inferringmethodspec,
author = {Pandita, Rahul and Xiao, Xusheng and Zhong, Hao and Xie, Tao and Oney, Stephen and Paradkar, Amit},
title = {Inferring method specifications from natural language API descriptions},
year = {2012},
isbn = {9781467310673},
publisher = {IEEE Press},
booktitle = {Proceedings of the 34th International Conference on Software Engineering},
pages = {815–825},
numpages = {11},
location = {Zurich, Switzerland},
series = {ICSE '12}
}

@inproceedings{randoop,
author = {Pacheco, Carlos and Ernst, Michael D.},
title = {Randoop: feedback-directed random testing for Java},
year = {2007},
isbn = {9781595938657},
publisher = {Association for Computing Machinery},
address = {New York, NY, USA},
url = {https://doi.org/10.1145/1297846.1297902},
doi = {10.1145/1297846.1297902},
booktitle = {Companion to the 22nd ACM SIGPLAN Conference on Object-Oriented Programming Systems and Applications Companion},
pages = {815–816},
numpages = {2},
location = {Montreal, Quebec, Canada},
series = {OOPSLA '07}
}

@misc{java21arraylist,
  author       = {{Oracle}},
  title        = {{ArrayList} ({Java SE 21 \& JDK 21 API Specification})},
  year         = {2023},
  howpublished = {\url{https://docs.oracle.com/en/java/javase/21/docs/api/java.base/java/util/ArrayList.html}},
  note         = {Accessed: 2026-06-30}
}

@misc{zheng2023judgingllmasajudgemtbenchchatbot,
      title={Judging LLM-as-a-Judge with MT-Bench and Chatbot Arena}, 
      author={Lianmin Zheng and Wei-Lin Chiang and Ying Sheng and Siyuan Zhuang and Zhanghao Wu and Yonghao Zhuang and Zi Lin and Zhuohan Li and Dacheng Li and Eric P. Xing and Hao Zhang and Joseph E. Gonzalez and Ion Stoica},
      year={2023},
      eprint={2306.05685},
      archivePrefix={arXiv},
      primaryClass={cs.CL},
      url={https://arxiv.org/abs/2306.05685}, 
}

@article{shriram,
   title={Little Tricky Logic: Misconceptions in the Understanding of LTL},
   volume={7},
   ISSN={2473-7321},
   url={http://dx.doi.org/10.22152/programming-journal.org/2023/7/7},
   DOI={10.22152/programming-journal.org/2023/7/7},
   number={2},
   journal={The Art, Science, and Engineering of Programming},
   publisher={Aspect-Oriented Software Association (AOSA)},
   author={Greenman, Ben and Saarinen, Sam and Nelson, Tim and Krishnamurthi, Shriram},
   year={2022},
   month=Oct 
}

@misc{bowman2015largeannotatedcorpuslearning,
      title={A large annotated corpus for learning natural language inference}, 
      author={Samuel R. Bowman and Gabor Angeli and Christopher Potts and Christopher D. Manning},
      year={2015},
      eprint={1508.05326},
      archivePrefix={arXiv},
      primaryClass={cs.CL},
      url={https://arxiv.org/abs/1508.05326}, 
}

@misc{fazelnia2024lessonsusenaturallanguage,
      title={Lessons from the Use of Natural Language Inference (NLI) in Requirements Engineering Tasks}, 
      author={Mohamad Fazelnia and Viktoria Koscinski and Spencer Herzog and Mehdi Mirakhorli},
      year={2024},
      eprint={2405.05135},
      archivePrefix={arXiv},
      primaryClass={cs.SE},
      url={https://arxiv.org/abs/2405.05135}, 
}

@inproceedings{MITCHELL1977VERSIONSPACES,
author = {Mitchell, Tom M.},
title = {Version spaces: a candidate elimination approach to rule learning},
year = {1977},
publisher = {Morgan Kaufmann Publishers Inc.},
address = {San Francisco, CA, USA},
booktitle = {Proceedings of the 5th International Joint Conference on Artificial Intelligence - Volume 1},
pages = {305–310},
numpages = {6},
location = {Cambridge, USA},
series = {IJCAI'77}
}

@misc{anthropic2026claudeopus48,
  author       = {{Anthropic}},
  title        = {Claude Opus 4.8},
  year         = {2026},
  howpublished = {\url{https://www.anthropic.com/claude/opus}},
  note         = {Accessed: 2026-06-30}
}

@misc{openai2026gpt55,
  author       = {{OpenAI}},
  title        = {Introducing {GPT-5.5}},
  year         = {2026},
  howpublished = {\url{https://openai.com/index/introducing-gpt-5-5/}},
  note         = {Accessed: 2026-06-30}
}

@article{hui2024qwen2,
      title={Qwen2. 5-Coder Technical Report},
      author={Hui, Binyuan and Yang, Jian and Cui, Zeyu and Yang, Jiaxi and Liu, Dayiheng and Zhang, Lei and Liu, Tianyu and Zhang, Jiajun and Yu, Bowen and Dang, Kai and others},
      journal={arXiv preprint arXiv:2409.12186},
      year={2024}
}

@misc{qwen3technicalreport,
      title={Qwen3 Technical Report}, 
      author={Qwen Team},
      year={2025},
      eprint={2505.09388},
      archivePrefix={arXiv},
      primaryClass={cs.CL},
      url={https://arxiv.org/abs/2505.09388}, 
}

@misc{openai2025gptoss120bgptoss20bmodel,
      title={gpt-oss-120b \& gpt-oss-20b Model Card}, 
      author={OpenAI},
      year={2025},
      eprint={2508.10925},
      archivePrefix={arXiv},
      primaryClass={cs.CL},
      url={https://arxiv.org/abs/2508.10925}, 
}

@article{Dillig-PLDI26ActiveLearning,
author = {Barnaby, Celeste and Ding, Danny and Bastani, Osbert and Dillig, I\c{s}\i{}l},
title = {Choose, Don't Label: Multiple-Choice Query Synthesis for Program Disambiguation},
year = {2026},
issue_date = {June 2026},
publisher = {Association for Computing Machinery},
address = {New York, NY, USA},
volume = {10},
number = {PLDI},
url = {https://doi.org/10.1145/3808279},
doi = {10.1145/3808279},
journal = {Proc. ACM Program. Lang.},
month = jun,
articleno = {201},
numpages = {25}
}

@article{ernst1999dynamically,
author = {Ernst, Michael D. and Cockrell, Jake and Griswold, William G. and Notkin, David},
title = {Dynamically Discovering Likely Program Invariants to Support Program Evolution},
year = {2001},
issue_date = {February 2001},
publisher = {IEEE Press},
volume = {27},
number = {2},
issn = {0098-5589},
url = {https://doi.org/10.1109/32.908957},
doi = {10.1109/32.908957},
journal = {IEEE Trans. Softw. Eng.},
month = feb,
pages = {99–123},
numpages = {25}
}

@article{ernst2007daikon,
   author = {Michael D. Ernst and Jeff H. Perkins and Philip J. Guo and
	Stephen McCamant and Carlos Pacheco and Matthew S. Tschantz and
	Chen Xiao},
   title = {The {Daikon} system for dynamic detection of likely invariants},
   journal = {Science of Computer Programming},
   volume = {69},
   number = {1--3},
   pages = {35--45},
   month = dec,
   year = {2007}
}

@inproceedings{ammons2002mining,
author = {Ammons, Glenn and Bod\'{\i}k, Rastislav and Larus, James R.},
title = {Mining specifications},
year = {2002},
isbn = {1581134509},
publisher = {Association for Computing Machinery},
address = {New York, NY, USA},
url = {https://doi.org/10.1145/503272.503275},
doi = {10.1145/503272.503275},
booktitle = {Proceedings of the 29th ACM SIGPLAN-SIGACT Symposium on Principles of Programming Languages},
pages = {4–16},
numpages = {13},
location = {Portland, Oregon},
series = {POPL '02}
}

@inproceedings{whaley2002automatic,
author = {Whaley, John and Martin, Michael C. and Lam, Monica S.},
title = {Automatic extraction of object-oriented component interfaces},
year = {2002},
isbn = {1581135629},
publisher = {Association for Computing Machinery},
address = {New York, NY, USA},
url = {https://doi.org/10.1145/566172.566212},
doi = {10.1145/566172.566212},
booktitle = {Proceedings of the 2002 ACM SIGSOFT International Symposium on Software Testing and Analysis},
pages = {218–228},
numpages = {11},
location = {Roma, Italy},
series = {ISSTA '02}
}

@inproceedings{xie2006automatic,
author = {Xie, Tao and Martin, Evan and Yuan, Hai},
title = {Automatic extraction of abstract-object-state machines from unit-test executions},
year = {2006},
isbn = {1595933751},
publisher = {Association for Computing Machinery},
address = {New York, NY, USA},
url = {https://doi.org/10.1145/1134285.1134427},
doi = {10.1145/1134285.1134427},
booktitle = {Proceedings of the 28th International Conference on Software Engineering},
pages = {835–838},
numpages = {4},
location = {Shanghai, China},
series = {ICSE '06}
}

@article{alur2005synthesis,
author = {Alur, Rajeev and \v{C}ern\'{y}, Pavol and Madhusudan, P. and Nam, Wonhong},
title = {Synthesis of interface specifications for Java classes},
year = {2005},
issue_date = {January 2005},
publisher = {Association for Computing Machinery},
address = {New York, NY, USA},
volume = {40},
number = {1},
issn = {0362-1340},
url = {https://doi.org/10.1145/1047659.1040314},
doi = {10.1145/1047659.1040314},
journal = {SIGPLAN Not.},
month = jan,
pages = {98–109},
numpages = {12}
}

@inproceedings{sankaranarayanan2008dynamic,
author = {Sankaranarayanan, Sriram and Chaudhuri, Swarat and Ivan\v{c}i\'{c}, Franjo and Gupta, Aarti},
title = {Dynamic inference of likely data preconditions over predicates by tree learning},
year = {2008},
isbn = {9781605580500},
publisher = {Association for Computing Machinery},
address = {New York, NY, USA},
url = {https://doi.org/10.1145/1390630.1390666},
doi = {10.1145/1390630.1390666},
booktitle = {Proceedings of the 2008 International Symposium on Software Testing and Analysis},
pages = {295–306},
numpages = {12},
location = {Seattle, WA, USA},
series = {ISSTA '08}
}

@inproceedings{padhi2016data,
author = {Padhi, Saswat and Sharma, Rahul and Millstein, Todd},
title = {Data-driven precondition inference with learned features},
year = {2016},
isbn = {9781450342612},
publisher = {Association for Computing Machinery},
address = {New York, NY, USA},
url = {https://doi.org/10.1145/2908080.2908099},
doi = {10.1145/2908080.2908099},
booktitle = {Proceedings of the 37th ACM SIGPLAN Conference on Programming Language Design and Implementation},
pages = {42–56},
numpages = {15},
location = {Santa Barbara, CA, USA},
series = {PLDI '16}
}

@inproceedings{astorga2019learning,
author = {Astorga, Angello and Madhusudan, P. and Saha, Shambwaditya and Wang, Shiyu and Xie, Tao},
title = {Learning stateful preconditions modulo a test generator},
year = {2019},
isbn = {9781450367127},
publisher = {Association for Computing Machinery},
address = {New York, NY, USA},
url = {https://doi.org/10.1145/3314221.3314641},
doi = {10.1145/3314221.3314641},
booktitle = {Proceedings of the 40th ACM SIGPLAN Conference on Programming Language Design and Implementation},
pages = {775–787},
numpages = {13},
location = {Phoenix, AZ, USA},
series = {PLDI 2019}
}

@article{astorga2021synthesizing,
author = {Astorga, Angello and Saha, Shambwaditya and Dinkins, Ahmad and Wang, Felicia and Madhusudan, P. and Xie, Tao},
title = {Synthesizing contracts correct modulo a test generator},
year = {2021},
issue_date = {October 2021},
publisher = {Association for Computing Machinery},
address = {New York, NY, USA},
volume = {5},
number = {OOPSLA},
url = {https://doi.org/10.1145/3485481},
doi = {10.1145/3485481},
journal = {Proc. ACM Program. Lang.},
month = oct,
articleno = {104},
numpages = {27}
}

@inproceedings{cosler2023nl2spec,
author = {Cosler, Matthias and Hahn, Christopher and Mendoza, Daniel and Schmitt, Frederik and Trippel, Caroline},
title = {nl2spec: Interactively Translating Unstructured Natural Language to Temporal Logics with Large Language Models},
year = {2023},
isbn = {978-3-031-37702-0},
publisher = {Springer-Verlag},
address = {Berlin, Heidelberg},
url = {https://doi.org/10.1007/978-3-031-37703-7_18},
doi = {10.1007/978-3-031-37703-7_18},
booktitle = {Computer Aided Verification: 35th International Conference, CAV 2023, Paris, France, July 17–22, 2023, Proceedings, Part II},
pages = {383–396},
numpages = {14},
location = {Paris, France}
}

@inproceedings{yan2024assertllm,
author = {Yan, Zhiyuan and Fang, Wenji and Li, Mengming and Li, Min and Liu, Shang and Xie, Zhiyao and Zhang, Hongce},
title = {AssertLLM: Generating Hardware Verification Assertions from Design Specifications via Multi-LLMs},
year = {2025},
isbn = {9798400706356},
publisher = {Association for Computing Machinery},
address = {New York, NY, USA},
url = {https://doi.org/10.1145/3658617.3697756},
doi = {10.1145/3658617.3697756},
booktitle = {Proceedings of the 30th Asia and South Pacific Design Automation Conference},
pages = {614–621},
numpages = {8},
location = {Tokyo, Japan},
series = {ASPDAC '25}
}

@misc{xia2026doc2spec,
      title={Doc2Spec: Synthesizing Formal Programming Specifications from Natural Language via Grammar Induction}, 
      author={Shihao Xia and Mengting He and Haomin Jia and Linhai Song},
      year={2026},
      eprint={2602.04892},
      archivePrefix={arXiv},
      primaryClass={cs.PL},
      url={https://arxiv.org/abs/2602.04892}, 
}

@misc{richter2025nl2contract,
      title={Beyond Postconditions: Can Large Language Models infer Formal Contracts for Automatic Software Verification?}, 
      author={Cedric Richter and Heike Wehrheim},
      year={2025},
      eprint={2510.12702},
      archivePrefix={arXiv},
      primaryClass={cs.SE},
      url={https://arxiv.org/abs/2510.12702}, 
}

@misc{specsyn,
      title={SpecSyn: LLM-based Synthesis and Refinement of Formal Specifications for Real-world Program Verification}, 
      author={Lezhi Ma and Shangqing Liu and Yi Li and Qiong Wu and Han Wang and Lei Bu},
      year={2026},
      eprint={2604.21570},
      archivePrefix={arXiv},
      primaryClass={cs.SE},
      url={https://arxiv.org/abs/2604.21570}, 
}

@misc{sldspec,
      title={Enhancing LLM-based Specification Generation via Program Slicing and Logical Deletion}, 
      author={Zehan Chen and Long Zhang and Zhiwei Zhang and JingJing Zhang and Ruoyu Zhou and Yulong Shen and JianFeng Ma and Lin Yang},
      year={2026},
      eprint={2509.09917},
      archivePrefix={arXiv},
      primaryClass={cs.SE},
      url={https://arxiv.org/abs/2509.09917}, 
}

@misc{autorespec,
      title={AutoReSpec: A Framework for Generating Specification using Large Language Models}, 
      author={Ragib Shahariar Ayon and Shibbir Ahmed},
      year={2026},
      eprint={2604.03758},
      archivePrefix={arXiv},
      primaryClass={cs.SE},
      url={https://arxiv.org/abs/2604.03758}, 
}

@misc{sun2024clover,
      title={Clover: Closed-Loop Verifiable Code Generation}, 
      author={Chuyue Sun and Ying Sheng and Oded Padon and Clark Barrett},
      year={2024},
      eprint={2310.17807},
      archivePrefix={arXiv},
      primaryClass={cs.AI},
      url={https://arxiv.org/abs/2310.17807}, 
}

@misc{claimcheck,
  author       = {Fernanda Graciolli and Nada Amin},
  title        = {{claimcheck}: Narrowing the Gap between Proof and Intent},
  howpublished = {\url{https://midspiral.com/blog/claimcheck-narrowing-the-gap-between-proof-and-intent/}},
  year         = {2026},
  month        = feb,
  note         = {Accessed: 2026-06-30}
}

@inproceedings{somers2005roundtrip,
    title = "Round-trip Translation: What Is It Good For?",
    author = "Somers, Harold",
    editor = "Baldwin, Timothy  and
      Curran, James  and
      van Zaanen, Menno",
    booktitle = "Proceedings of the Australasian Language Technology Workshop 2005",
    month = dec,
    year = "2005",
    address = "Sydney, Australia",
    url = "https://aclanthology.org/U05-1019/",
    pages = "127--133"
}

@misc{uppar2026verificationmodulotestedlibrary,
      title={Verification Modulo Tested Library Contracts}, 
      author={Abhishek Uppar and Omar Muhammad and Sumanth Prabhu and Deepak D'Souza and Madhusudan P and Adithya Murali},
      year={2026},
      eprint={2604.15533},
      archivePrefix={arXiv},
      primaryClass={cs.PL},
      url={https://arxiv.org/abs/2604.15533}, 
}

\appendix

~\\
\newpage
\section{Detailed Statistics of Datasets}
\label{app:dataset}

Table~\ref{tab:dataset-composition} reports the detailed composition of the datasets used in our evaluation. Each entry counts the number of assertion-generation data points for a given class, where one data point consists of a code context, a natural-language specification, and the corresponding ground-truth formal assertion. The C2S-Augmented dataset is our main benchmark and contains 416 data points across 19 Java collection-style classes. C2S-Augmented Sub is a 96-data-point subset used for more expensive ablation and comparison experiments. Manual NL contains 39 human-written natural-language specifications over three additional classes. The Buggy Code and Buggy Assertion datasets are derived from selected C2S-Augmented classes and are used to evaluate robustness under perturbed implementations and incorrect assertions, respectively.

\begin{table*}[t]
\centering
\footnotesize
\setlength{\tabcolsep}{6pt}
\renewcommand{\arraystretch}{1.08}
\begin{tabular}{lrrrrr}
\toprule
\textbf{Class} & \textbf{C2S-Aug.} & \textbf{C2S-Aug. Sub} & \textbf{Manual NL} & \textbf{Buggy Code} & \textbf{Buggy Assertion} \\
\midrule
ArrayDeque             & 30 & 8  & -- & 3 & 8  \\
ArrayList              & 32 & 17 & -- & 4 & 15 \\
ArrayStack             & 5  & -- & -- & -- & -- \\
AttributeList          & 16 & -- & -- & -- & -- \\
BitSet                 & 49 & 30 & -- & 0 & 0  \\
CircularFifoQueue      & 15 & -- & -- & -- & -- \\
DefaultListenableGraph & 4  & 4  & -- & 0 & 0  \\
FixedArrayList         & 6  & -- & -- & -- & -- \\
HashMap                & 20 & 12 & -- & 0 & 0  \\
HashSet                & 16 & 5  & -- & 2 & 5  \\
LinkedList             & 61 & -- & -- & -- & -- \\
ListOrderedSet         & 9  & -- & -- & -- & -- \\
Path                   & 6  & -- & -- & -- & -- \\
PriorityQueue          & 24 & 18 & -- & 4 & 13 \\
RoleList               & 26 & -- & -- & -- & -- \\
RoleUnresolvedList     & 26 & -- & -- & -- & -- \\
Stack                  & 5  & 2  & -- & 2 & 3  \\
TreeList               & 20 & -- & -- & -- & -- \\
Vector                 & 46 & -- & -- & 5 & 22 \\
TreeSet                & -- & -- & 13 & -- & -- \\
Trie                   & -- & -- & 17 & -- & -- \\
UnionFind              & -- & -- & 9  & -- & -- \\
\midrule
\textbf{Total}         & \textbf{416} & \textbf{96} & \textbf{39} & \textbf{20} & \textbf{66} \\
\bottomrule
\end{tabular}
\caption{Dataset composition by class. Each entry reports the number of assertion-generation data points. ``--'' indicates that the class is not included in the dataset.}
\label{tab:dataset-composition}
\end{table*}

\section{Prompt Design}
\label{app:prompt}

\subsection*{Candidate-generation prompt}

The candidate-generation prompt asks the LLM to translate the natural-language assertion marked by \texttt{@@@} into exactly one executable assertion. The prompt includes a small number of in-context examples and explicitly describes the supported logic syntax, including \texttt{\textbackslash old(...)} for pre-state values, \texttt{\textbackslash result} for return values, \texttt{=>} for implication, and \texttt{\textbackslash forall} for universal quantification.

\begin{lstlisting}[basicstyle=\ttfamily\footnotesize,breaklines=true,columns=fullflexible]
Your task is to read java code and output exactly one assert statement (specification) corresponding to the comment that starts with "@@@". 
Your statement should accurately reflect the natural language assertion. Make sure to include all relevant details (such as premises, bounds, or branch cases), and avoid leaving out information or adding anything extraneous.
You can think step by step before coming up with the final answer, but please make sure that your final answer (the java code of the assert statement) is displayed in a <code></code> block.

Here are some rules and tips:
1. If the assert statement needs to access the value of an expression at the beginning of the function, you can use the `\old(expression)` syntax.
2. To refer to the return value of the method in the assert statement, you can use the `\result` variable.
3. To write A implies B or B happens if A holds, you may use the `A => B` syntax. Keep in mind that `=>` is binary with exactly two operands, and `=>` has lower priority than operator `&&` and `||` so use parentheses to ensure correct grouping.
4. If the assert statement needs to express that the `spec` should hold for each int variable `i` where `cond` holds, you may use `\forall int i; cond; spec` syntax.
5. Use only publicly accessible observer methods provided in the test function. (Use `\result` instead of calling the observer method that is being tested.) All function calls in the test function should be of the format `this.func_name(args_list)`.

[Few-shot Java examples]

Now, read the following java code and output an assert statement corresponding to the comment that starts with "@@@".

{TARGET_CODE}
\end{lstlisting}

\subsection*{Roundtrip translation prompt}
For roundtrip checking, we first translate a generated formal assertion back into natural language. We use the precise mode so that the roundtripped description preserves premises, bounds, old-state references, return-value references, null cases, equality conditions, and logical connectives.

\begin{lstlisting}[basicstyle=\ttfamily\footnotesize,breaklines=true,columns=fullflexible]
Your task is to read java code and write a natural language assertion that describes a specific logical specification in the code.

Here are descriptions for some syntactic symbols in JML logic:
1. If the assertion uses "=>" or "==>", such as "A => B", that means "A" implies "B".
2. \old(expression) means the value of expression before the method is executed.
3. \result means the return value of the method.
4. "\forall var i; cond; spec" means that the "spec" should hold for all the "i" such that "cond" holds.

Please try to make your translation as consistent as possible. Your translation should be equivalent to the original assertion. Do not include anything extra, and also do not leave anything out, especially premises.
Also write the translation as if you were a careful human engineer writing contract specifications for this code in plain English: follow the logic of the code naturally, make the statement fluent, clear, and easy to read, and prefer ordinary human phrasing over formal logical wording.

Translate the formal assertion as accurately and completely as possible: preserve every premise, bound, quantified condition, old-state reference, return-value reference, null case, equality condition, and logical connective in the natural language assertion.
Do not summarize away implementation-level details when they affect the exact meaning. Include all details from the formal formula in natural language, even if the resulting sentence is longer.
For example, if an assertion says something like "index>=0 && index<=size", explicitly state that the index is greater than or equal to 0 and less than or equal to size. Likewise, if an assertion uses a null-safe equality pattern like "a==null && b==null || a.equals(b)", explicitly mention both the case where both values are null and the case where one value equals the other.

Your output should be one single line starting with "// @@@ "
Here are several examples for your reference:

[Few-shot Java examples]

Now, read the following java code:

{TARGET_CODE}

Write a natural language assertion that describes the assertion code: {TARGET_ASSERTION}.
Please directly output your answer without any other information.

\end{lstlisting}

\subsection*{Conformance-scoring prompt}
After roundtrip translation, we compare the original natural-language assertion with the roundtripped natural-language assertion. Our main conformance checker uses a contextual, component-level prompt. It asks the model to score how well the hypothesis preserves the premise, where the premise is the original specification and the hypothesis is the roundtripped specification.

\begin{lstlisting}[basicstyle=\ttfamily\footnotesize,breaklines=true,columns=fullflexible]
You are performing a **roundtrip conformance check** between:
(1) a natural-language post-condition (the "premise"), and
(2) a second natural-language statement (the "hypothesis") that is a natural-language translation of a generated post-condition formula.

Your job is to judge how well the hypothesis conforms to the premise under the following responsibility:

Conformance responsibility
1) Completeness: The hypothesis must not omit any component that the premise intends.
   - Every component mentioned or implied as an intended constraint in the premise should be present in, or entailed by, the hypothesis.
2) Soundness: The hypothesis should not introduce unrelated or unjustified components.
   - Ideally, every component in the hypothesis should be traceable to the premise's intention.
   - In practice, *minor reasonable supplementation* is allowed (e.g., explicit bounds/ranges, type conversions, edge-case handling) as long as it does not change the intended structure or meaning.

What counts as a "component" (treat these as the primary comparison units)
- Entities / terms: the return value or result, method parameters, object fields, collection elements, indices, sizes, old/pre-state values, and new/post-state values
- Quantification / scope: statements about all items, any item, no item, a particular item, or a restricted subset of items
- Bounds / ranges: valid index ranges, quantified ranges, numeric limits, and whether endpoints are included or excluded
- Predicate relations: equality, inequality, ordering/comparison, containment/membership, nullness, type/compatibility requirements, and method-call properties
- Logical structure: conditions ("if/when/only if"), conjunctions ("and"), alternatives ("or/either"), negation ("not/no"), implications, case splits, and grouping/precedence implied by the sentence

Evaluation principle
- This is a *structure- and alignment-focused* check: prioritize whether the hypothesis preserves the overall logical skeleton and aligns each segment/component to the premise.
- Ignore superficial wording differences and synonyms of technical terms.
- Ignore fine-grained details; focus on whether the same structural components are present and aligned, and whether their logical relationships (quantifiers, bounds, predicates, connectives) match.
- The hypothesis may be imperfectly phrased or not fully "compilable" as a formula; still score based on whether the intended components/structure match.
- Do not purely evaluate on literal text similarity; instead, reason about semantics and structure in the context of target code behavior. (The context information is given below.)

Scoring (real value in [-1.0, 1.0])
Interpret the score as a combined measure of:
- Completeness (missing components/segments -> lower score)
- Soundness (unjustified extra components/segments -> lower score)
- Logical compatibility (contradictions/incompatible constraints -> negative score)

Use these anchor points (you may output intermediate values like 0.8, 0.2, -0.3):
+1.0 Perfect Conformance:
  - Hypothesis preserves all components and the logical structure of the premise (may rephrase wording).
  - Hypothesis may add minor, clearly reasonable supplementation (e.g., explicit bounds/ranges, type conversions, edge-case handling) as long as it does not change the intended structure or meaning.
+0.5 Mostly Conformant / Minor Loss:
  - Hypothesis is consistent with the premise but omits some non-trivial components, weakens constraints, or only covers a subset of cases.
  - Or adds some unrelated or unjustified content while the core structure still mostly aligns with the premise.
0.0 Neutral / Not Established:
  - The hypothesis is largely unrelated to the premise's components/structure, OR
  - The hypothesis introduces new requirements not supported by the premise, OR
  - The hypothesis omits essential segments such that conformance cannot be verified from it.
-0.5 Partially Conflicting:
  - Some aligned components exist, but at least one important segment contradicts the premise (e.g., flipped inequality, negation, wrong old/pre-state reference, implication reversed, incompatible bound).
-1.0 Completely Conflicting:
  - The core structure/meaning contradicts the premise and cannot hold simultaneously.

For instance,
[Few-shot examples]

Here is the context information to help understand the sentences:
The sentences are assertions in natural language form describing the expected behavior or properties of a piece of java code, such as a function or method.
More specifically, you are given the implementation of a class [Class name]. Inside this class, there is a method [Method name]. The following two assertions are both written about the behavior of the method. To help you reason about their relationship, here is some context:
- The method takes in ([Parameters]) as parameters.
- [Return information]
- The functionality of the method is documented below:
[Method documentation]

Output format:
- First, briefly explain your reasoning process in 1-3 concise sentences.
- Then, output only your decision as real-valued score wrapped in <ans> </ans> tags.
- Place exactly one score inside the tags. Do not include anything else inside the tags.

Here are the two assertions to evaluate:
The premise is: [Premise]
The hypothesis is: [Hypothesis]
\end{lstlisting}

\section{Details on Test Harness Construction}
\label{app:test-code}

For each generated assertion, our implementation constructs an executable Java test harness that checks the assertion over inputs generated by Randoop. The construction has four main steps: parsing the target context, translating the assertion formula, composing the target-method call, and embedding the translated formula into a test method.

\paragraph{Parsing the target context}
Given the code skeleton for a data point, the implementation first extracts the class name, import declarations, method signatures, method arguments, return type, documentation string, and whether the target method is a constructor. We treat the last method in the skeleton as the target method to be checked. The remaining public observer methods, such as \texttt{size()}, \texttt{get()}, \texttt{contains()}, \texttt{indexOf()}, and \texttt{isEmpty()}, are available to generated assertions and to the generated test code. The generated test method takes an instance of the target class, named \texttt{fuzzobj}, together with the original target-method arguments. For example, for \texttt{ArrayList.add(int index, E element)}, the test method receives \texttt{ArrayList<E> fuzzobj}, \texttt{int index}, and \texttt{E element}.

\paragraph{Extracting the assertion formula}
The checker expects each candidate to contain exactly one assertion of the form \texttt{assert formula;}. The implementation extracts the inner formula and normalizes implication operators such as \texttt{==>}, \texttt{->}, and \texttt{=>} into a single implication operator \texttt{=>}. It then rewrites references to the receiver and return value: \texttt{this} is replaced with the generated receiver variable \texttt{fuzzobj}, and \texttt{\textbackslash result} is replaced with the generated return variable \texttt{New\_Ret}. The resulting formula is the basis for all later code generation.

\paragraph{Handling pre-state expressions}
Assertions may refer to pre-state values using \texttt{\textbackslash old(e)}. For each such expression, the implementation creates a fresh variable, such as \texttt{OLD\_var0}, and emits code that evaluates the expression before the target method is executed:
\begin{lstlisting}[language=Java,basicstyle=\ttfamily\footnotesize]
var OLD_var0 = exec(() -> fuzzobj.size());
\end{lstlisting}
The occurrence of \texttt{\textbackslash old(fuzzobj.size())} in the assertion is then replaced by \texttt{OLD\_var0}. Nested \texttt{\textbackslash old} expressions are handled recursively, so that all required pre-state values are saved before any mutation performed by the target method.

\paragraph{Handling quantified assertions}
The simplified assertion language supports integer universal quantification of the form:
\begin{lstlisting}[basicstyle=\ttfamily\footnotesize]
\forall int i; condition; specification
\end{lstlisting}
The implementation extracts the quantified variable, the range condition, and the quantified body. It then infers loop bounds from simple inequalities in the condition. For example, a condition such as \texttt{0 <= i \&\& i < oldSize} is converted into a loop over the corresponding integer range. The quantified formula is translated into a Boolean accumulator:
\begin{lstlisting}[language=Java,basicstyle=\ttfamily\footnotesize]
Boolean forall_holds = Boolean.TRUE.equals(exec(() -> {
  boolean ret = true;
  for (int i = lo; i <= hi; i++) {
    if (condition) {
      ret &= Boolean.TRUE.equals(exec(() -> specification));
    }
  }
  return ret;
}));
\end{lstlisting}
The original \texttt{\textbackslash forall} expression is then replaced by the generated Boolean variable. If the quantified body contains \texttt{\textbackslash old} expressions that depend on the quantified variable, the implementation snapshots the corresponding pre-state values into an array before the method call and retrieves them during the loop.

\paragraph{Splitting compound formulas}
To make assertion evaluation robust and to localize failures, compound formulas are split into intermediate expressions. The implementation recursively decomposes formulas at top-level logical operators, including implication, disjunction, conjunction, and equality. Implication \texttt{A => B} is translated into \texttt{!A || B}. Equality between non-null expressions is translated using \texttt{Objects.equals}, while equality against \texttt{null} is translated using Java reference equality. Each intermediate expression is evaluated through \texttt{exec}, which catches runtime exceptions and converts them to \texttt{null}. Boolean subformulas are then normalized with \texttt{Boolean.TRUE.equals(...)} so that exceptions and non-true results are treated as failures.

\paragraph{Composing the target-method call}
After all pre-state values have been copied, the test harness executes the target method. Void methods are invoked through a \texttt{Runnable} wrapper:
\begin{lstlisting}[language=Java,basicstyle=\ttfamily\footnotesize]
String exceptionType =
  func_call_runnable(() -> fuzzobj.add(index, element));
\end{lstlisting}
Methods with return values are invoked through a \texttt{Supplier} wrapper, which stores both the return value and any thrown exception:
\begin{lstlisting}[language=Java,basicstyle=\ttfamily\footnotesize]
var paired_ret =
  func_call_supplier(() -> fuzzobj.method(args));
var New_Ret = paired_ret.first;
String exceptionType = paired_ret.second;
\end{lstlisting}
Constructors are handled similarly, except that the constructed object is stored as the post-state receiver. This uniform wrapping lets the generated test distinguish exceptions raised by the target method from exceptions raised while evaluating the generated assertion.

\paragraph{Checking the postcondition}
After the target method has executed, the generated code evaluates the translated assertion formula. The normal postcondition is checked as follows:
\begin{lstlisting}[language=Java,basicstyle=\ttfamily\footnotesize]
Boolean normalpost = exec(() -> translatedFormula);
if (normalpost == null || !normalpost) {
  throw new RuntimeException("Normal Postcondition Violated");
}
\end{lstlisting}
Thus, a candidate fails if the assertion evaluates to \texttt{false} or if evaluating the assertion raises an exception. The latter is important because many incorrect generated assertions are syntactically plausible but unsafe, for example by calling \texttt{equals} on a value that may be \texttt{null}. Exceptional postconditions are currently represented by a default true condition unless the checker is configured with additional exception-specific constraints.

\begin{figure}[t]
\centering
\begin{minipage}{0.98\columnwidth}
\begin{lstlisting}[language=Java,basicstyle=\ttfamily\footnotesize]
public class FuzzTest<E> {
  public void testAdd(ArrayList<E> list, int index, E element) {
    if (list == null) return;

    // copy old values
    Integer oldSize = exec(() -> list.size());

    // method execution
    String ex = funcCall(() -> list.add(index, element));

    // candidate assertion:
    Boolean pre = exec(() -> index >= 0 && index <= oldSize);
    Boolean post = exec(() -> list.get(index).equals(element));
    Boolean assertion = exec(() -> !pre || post);

    if (assertion == null || !assertion) {
      throw new RuntimeException("Postcondition Violated");
    }
  }

  private static <T> T exec(Supplier<T> s) {
    try {
      return s.get();
    } catch (Exception e) {
      return null;
    }
  }

  private static String funcCall(Runnable r) {
    try {
      r.run();
      return null;
    } catch (Exception e) {
      return e.getClass().getSimpleName();
    }
  }
}
\end{lstlisting}
\end{minipage}
\vspace{-0.8em}
\caption{Simplified fuzz test generated from the running example in Section~\ref{sec:example}.}
\label{fig:fuzz-test-example}
\end{figure}

Figure~\ref{fig:fuzz-test-example} shows a simplified fuzz test generated from the example in Section~\ref{sec:example}. The test first saves the pre-state value needed by the assertion, namely \texttt{list.size()}, to implement \texttt{\textbackslash old(this.size())}. It then executes the target method, \texttt{list.add(index, element)}, on inputs generated by the test generator. After the method call, the candidate assertion is evaluated in two parts: the precondition checks whether \texttt{index} is within the valid insertion range in the pre-state, and the postcondition checks whether \texttt{list.get(index).equals(element)} holds. The implication is encoded as \texttt{!pre || post}. If this expression evaluates to \texttt{false}, or if evaluating it raises an exception and therefore returns \texttt{null}, the test throws a runtime exception, causing the generated input to be reported as a counterexample. The helper function \texttt{exec} safely evaluates expressions that may throw exceptions during assertion checking. Similarly, \texttt{funcCall} executes the target method while recording whether the method itself throws an exception. In this way, the fuzz test distinguishes exceptions caused by the program under test from exceptions caused by an invalid or unsafe generated assertion.

\paragraph{Helper functions}
The generated harness includes a small set of helper functions to make assertion evaluation robust. We describe them below.

\textbf{Pair.}
For target methods with return values, the harness needs to record both the returned value and whether the method itself throws an exception. We use a lightweight \texttt{Pair} container for this purpose.

\begin{lstlisting}[language=Java,basicstyle=\ttfamily\footnotesize]
private static class Pair<A, B> {
  private final A first;
  private final B second;

  private Pair(A first, B second) {
    this.first = first;
    this.second = second;
  }
}
\end{lstlisting}

\textbf{Safe expression evaluation.}
Generated assertions may be unsafe even when they are syntactically valid; for example, they may call \texttt{equals} on a \texttt{null} value. The \texttt{exec} helper evaluates an assertion subexpression and converts any runtime exception into \texttt{null}. The checker then treats \texttt{null} as a failed assertion evaluation.

\begin{lstlisting}[language=Java,basicstyle=\ttfamily\footnotesize]
private static <T> T exec(Supplier<T> supplier) {
  try {
    return supplier.get();
  } catch (Exception fuzzexception) {
    return null;
  }
}
\end{lstlisting}

\textbf{Method execution.}
The harness separates execution of the target method from evaluation of the generated assertion. For target methods with return values, \texttt{func\_call\_supplier} stores both the returned value and any exception tag. For void methods, \texttt{func\_call\_runnable} records only whether the method throws. This separation allows the checker to distinguish exceptions raised by the program under test from exceptions raised later during assertion evaluation.

\begin{lstlisting}[language=Java,basicstyle=\ttfamily\footnotesize]
private static <T> Pair<T, String> func_call_supplier(Supplier<T> supplier) {
  try {
    return new Pair<>(supplier.get(), null);
  } catch (Exception fuzzexception) {
    String exceptionType = fuzzexception.getClass().getSimpleName();
    return new Pair<>(null, exceptionType);
  }
}

private static String func_call_runnable(Runnable runnable) {
  try {
    runnable.run();
    return null;
  } catch (Exception fuzzexception) {
    return fuzzexception.getClass().getSimpleName();
  }
}
\end{lstlisting}

\textbf{Retrieving saved pre-state values.}
For quantified assertions, a \texttt{\textbackslash old} expression may depend on the quantified variable. In such cases, the harness snapshots the relevant pre-state values into an array before executing the target method. During assertion evaluation, \texttt{get\_from\_array} retrieves the saved value at the current quantified index and casts it back to the expected target type.

\begin{lstlisting}[language=Java,basicstyle=\ttfamily\footnotesize]
@SuppressWarnings("unchecked")
private static <T> T get_from_array(Object arr, int index, T ex_val) {
  try {
    return (T) Array.get(arr, index);
  } catch (Exception fuzzexception) {
    return null;
  }
}
\end{lstlisting}

\paragraph{Compilation and fuzzing}
The composed test harness is written into a temporary \texttt{fuzztests.FuzzTest} class and compiled together with the target class and required third-party libraries. The compile check succeeds only if this generated harness compiles. The fuzz check then invokes Randoop on the generated \texttt{FuzzTest} class. If Randoop reports no error-revealing tests, the candidate passes the fuzz check; otherwise, the candidate is marked invalid. This design allows the same generated harness to serve both as a compilation oracle and as an executable semantic check for generated assertions.

\begin{figure*}[t]
\centering
\includegraphics[width=0.98\textwidth]{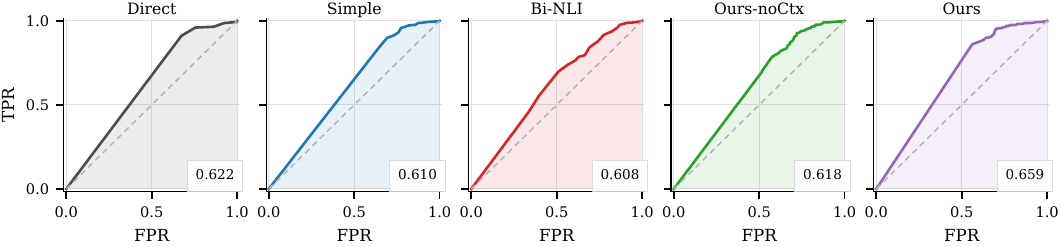}
\caption{Overall ROC curves for conformance implementations using all generated candidates across datasets. Dashed lines denote baseline methods; solid lines denote our clausal variants.}
\label{fig:conf-roc-overall}
\end{figure*}

\begin{figure*}[t]
\centering
\includegraphics[width=0.98\textwidth]{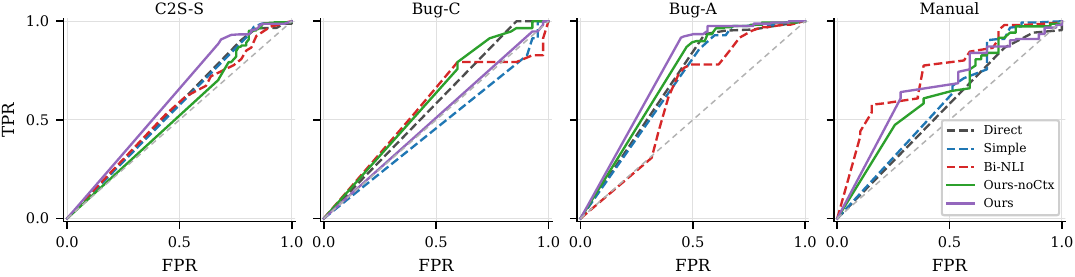}
\caption{Per-dataset ROC curves for conformance implementations. Each panel corresponds to one dataset, and each curve corresponds to one conformance implementation.}
\label{fig:conf-roc-by-dataset}
\end{figure*}

\paragraph{Equivalence-Check Harness}
The equivalence check reuses the same executable-harness infrastructure described above, but changes what is executed and what property is checked. The fuzz and validity checks are context-dependent: they execute the target method on generated inputs and then evaluate one candidate assertion against the resulting program state. In contrast, the equivalence check is context-independent. It does not call the target method and does not depend on whether a particular receiver object satisfies the original method semantics. Instead, it compares two assertion formulas directly over the same generated valuation of their free variables.

Concretely, given two formulas \(A\) and \(B\), the generated equivalence harness evaluates both formulas under the same generated inputs and checks whether they always agree:
\begin{lstlisting}[language=Java,basicstyle=\ttfamily\footnotesize]
Boolean left = exec(() -> translatedFormulaA);
Boolean right = exec(() -> translatedFormulaB);

Boolean equiv = exec(() -> Objects.equals(left, right));

if (equiv == null || !equiv) {
  throw new RuntimeException("Equivalence Violated");
}
\end{lstlisting}

Thus, Randoop is used to search for a valuation that distinguishes the two formulas. The main difference from the fuzz-check harness is therefore the absence of the target-method call. There is no pre-state/post-state transition to execute, and no candidate is checked against the implementation behavior of a method. Instead, both formulas are translated into executable Java expressions and evaluated side by side. This makes the equivalence check a formula-level test: it asks whether two candidate assertions describe the same condition over the generated inputs, independent of the method body or the correctness of the implementation.

\section{Experimental Setup}
\label{app:experimental-setup}

Our main experiments use two open-source backbone models: \textsc{Qwen2.5-Coder-32B-Instruct}~\cite{hui2024qwen2}, a smaller code-specialized model, and \textsc{GPT-oss-120B}~\cite{openai2025gptoss120bgptoss20bmodel}, a larger instruction-tuned model. For each natural-language specification, we sample five candidates with a moderately high temperature $0.6$ to encourage output diversity, and set the maximum generation length large enough ($65536$) to avoid truncating formal assertions. The experiments are run on a local server with two NVIDIA L40S GPUs.

\section{Detailed Explanation and Visualization for Different Conformance Implementation}
\label{app:conf-impl}

This appendix describes the conformance-checking variants evaluated in Section~\ref{sec:eval}. Each method takes as input the original natural-language specification and a generated formal assertion, and returns a conformance score estimating whether the generated assertion preserves the intended meaning of the specification. The variants differ in whether they use roundtrip translation, whether they decompose the specification into components, whether the comparison is bidirectional, and whether they include contextual examples.

\paragraph{Direct}
The Direct baseline does not use roundtrip translation. It directly asks the LLM whether the generated formal assertion is equivalent to the original natural-language specification. To make the formal assertion interpretable, the prompt first describes our JML-like assertion syntax, including implication, quantification, \texttt{\textbackslash old}, \texttt{\textbackslash result}, and receiver references such as \texttt{this}. It also provides the scoring rule used by the checker and the relevant code context, so that the model can interpret the formula with respect to the target method. The model then receives the natural-language intent and the candidate assertion, and predicts whether they describe the same behavior. Because the model must reason directly over the formal logic formula, this method depends heavily on the LLM's ability to understand the assertion syntax and map it to the intended semantics.

\paragraph{Simple}
Simple is a roundtrip-based baseline. It first asks the LLM to translate the generated formal assertion back into natural language. It then asks whether this generated natural-language description is equivalent to the original natural-language specification. This reduces the comparison to a simple natural-language equivalence judgment, but it does not explicitly ask to compare bidirectionally, or decompose the specification into smaller semantic parts.

\paragraph{Bi-NLI}
Bi-NLI uses roundtrip translation, but frames the comparison as a bidirectional natural-language inference task. After translating the formal assertion back into natural language, the checker asks whether the original specification entails the roundtripped description and whether the roundtripped description entails the original specification. Unlike our full method, this variant does not provide additional code context, examples, or component-level alignment guidance during the equivalence judgment. The candidate is considered conforming only when both entailment directions hold. This formulation makes equivalence explicit, but can be sensitive to conservative entailment judgments and score calibration.

\paragraph{Ours-noCtx}
Ours-noCtx is a clausal-level roundtrip method without additional context. It decomposes the original specification and the roundtripped description into semantic clauses, then checks whether the clauses align bidirectionally. This lets the checker compare finer-grained pieces of meaning rather than making a single global equivalence judgment. However, because this variant omits contextual examples and surrounding task information, the model has less guidance for interpreting edge cases or domain-specific intent.

\paragraph{Ours}
Ours is our full conformance implementation. LikeOurs-noCtx, it uses roundtrip translation and bidirectional clause-level alignment. In addition, it provides contextual information, including the target code context and examples of how specifications should be interpreted. This helps the LLM resolve ambiguous terms, align clauses more consistently, and avoid judging candidates only by surface-level textual similarity. This is the method used in the main pipeline unless otherwise stated.

\subsection*{Visualization}

Figure~\ref{fig:conf-roc-overall} compares the overall ROC curves of all conformance implementations after pooling candidates from all datasets. The curve for \textit{Ours} dominates the alternatives in aggregate, showing that the contextual bidirectional clausal checker provides the strongest ranking signal for distinguishing equivalent from non-equivalent candidates. Figure~\ref{fig:conf-roc-by-dataset} breaks the same comparison down by dataset. The per-dataset curves show that no single baseline is uniformly best across all settings, but \textit{Ours} is the most stable overall and obtains the highest aggregate AUC.

\end{document}